\documentclass[preprint]{elsarticle}

\date{May 13, 2020}
\usepackage{lineno,hyperref}
\modulolinenumbers[5]         

\journal{Information Systems}

\graphicspath{{./figs/}}

\usepackage{amssymb}
\usepackage{amsmath}
\usepackage{graphicx}
\usepackage{mathrsfs}
\usepackage{algorithmic}
\usepackage{subfig}
\usepackage{url}
\usepackage{multirow} % table
\usepackage{threeparttable} % table
\usepackage{xfrac} % better fraction display
\usepackage{algorithm} % use algorithm wrapper only with non-floating control
\usepackage{amsthm} % unnumbered theorem
\usepackage{xcolor} % highlighting purpose
\usepackage{lscape} % rotate large figures

\newcommand{\todo}{\textcolor{red}{\textbf{[TODO]}} } % use this to put comments for TODO work items
\newif\ifcomment \commentfalse % use this to comment out unwanted latex code

\newcommand{\mi}{\mathit} % for math
\newcommand{\powerset}{\mathcal{P}}
\newcommand{\bag}{\mathcal{B}}
\newcommand{\univ}{\mathcal{U}}
\newcommand{\card}[1]{\left\lvert{#1}\right\rvert}
\let\emptyset\varnothing % better looking \emptyset symbol: substitute it with \varnothing

\newtheorem*{method*}{Method}
\newtheorem{definition}{Definition}

\begin{document}

\begin{frontmatter}

\title{\emph{OrgMining~2.0}: A Novel Framework for\\ Organizational Model Mining from Event Logs} 

\author[QUT]{Jing~Yang\corref{correspondingauthor}}
\cortext[correspondingauthor]{Corresponding author}
\ead{roy.j.yang@qut.edu.au}

\author[QUT]{Chun~Ouyang}
\ead{c.ouyang@qut.edu.au}

\author[RWTH,QUT]{Wil~M.P.~van~der~Aalst}
\ead{wvdaalst@pads.rwth-aachen.de}

\author[QUT]{Arthur~H.M.~ter~Hofstede}
\ead{a.terhofstede@qut.edu.au}

\author[SYSU]{Yang~Yu}
\ead{yuy@mail.sysu.edu.cn}

\address[QUT]{Queensland University of Technology, Australia}
\address[RWTH]{RWTH Aachen University, Germany}
\address[SYSU]{Sun Yat-sen University, China}

\begin{abstract}
Providing appropriate structures around human resources can streamline operations and thus facilitate the competitiveness of an organization. 
To achieve this goal, modern organizations need to acquire an accurate and timely understanding of human resource grouping while faced with an ever-changing environment.
The use of process mining offers a promising way to help address the need through utilizing event log data stored in information systems. By extracting knowledge about the actual behavior of resources participating in business processes from event logs, organizational models can be constructed, which facilitate the analysis of the {de facto} grouping of human resources relevant to process execution.
Nevertheless, open research gaps remain to be addressed when applying the state-of-the-art process mining to analyze resource grouping.
For one, the discovery of organizational models has only limited connections with the context of process execution.
For another, a rigorous solution that evaluates organizational models against event log data is yet to be proposed.
In this paper, we aim to tackle these research challenges by developing a novel framework built upon a richer definition of organizational models coupling resource grouping with process execution knowledge. By introducing notions of conformance checking for organizational models, the framework allows effective evaluation of organizational models, and therefore provides a foundation for analyzing and improving resource grouping based on event logs.
We demonstrate the feasibility of this framework by proposing an approach underpinned by the framework for organizational model discovery, and also conduct experiments on real-life event logs to discover and evaluate organizational models.

\end{abstract}


% ------------------------------ Bullet points for "Highlights" ---------------------------------

% * Propose a novel definition of organizational model for process mining which couples human resource structure with business process

% * Build a framework allowing not only discovery but also conformance checking of organizational models using event log data

% * Establish a foundation for process mining research in terms of investigating the organizational perspective, especially the structure and behavior of resources

% ------------------------------ Bullet points for "Highlights" ---------------------------------

\end{frontmatter}

%\linenumbers               % comment this out for arXiv uploading

\section{Introduction}
\label{sec:intro}

% Organization: people, processes and changes
Modern organizations are comprised of employees working in various group settings, often directed by end-to-end business processes towards delivering valued outcomes to achieve the organizations' goals~\cite{daft2010organization}. Faced with an ever-changing environment, organizations today are required to constantly adapt their settings in order to respond to various and often dynamic demands. Having flexible and proper structures around human resources may facilitate addressing such a challenge, as evidenced by the organizational structure-related practices for business process redesign~\cite{reijers2005best}. For this purpose, organizations need to both obtain and maintain an accurate and timely understanding of human resource grouping alongside evolving business processes. 
This, however, can hardly be achieved by merely having a static ``as-is'' organizational chart.

% Information Systems, data and process mining

A promising way to address the above dynamic need is through utilizing data related to process execution readily available in many information systems (e.g., Enterprise Resource Planning systems) deployed in today's organizations~\cite{vanderaalst2016process}.
These data are often stored as event logs.
A typical event log often records some essential information on the execution of a business process, including the activities occurred, their timestamps, and the corresponding process instance (e.g., a purchase order).
Additionally, event logs may also contain human resource information, thus reflecting how employees carried out their work when participating in processes. 
In light of this, event logs may be leveraged for discovering knowledge about the structure, behavior, and performance of human resource groups, and the potential changes to human resource grouping in the context of business processes.

% Process mining for the purpose: organizational model mining
Process mining~\cite{vanderaalst2016process} offers a growing body of methods to analyze event logs to extract knowledge about the actual behavior of a process along with other aspects relevant to process execution, including human resources. In process mining research, studies that concern human resource grouping are known as \emph{organizational model mining}, which aims at finding groups of resources having similar characteristics in process execution by utilizing event log data~\cite{song2008towards}. 
Our research interest lies in applying organizational model mining to support the analysis and improvement of resource grouping based on event log data.
Nonetheless, our review of the literature on the topic reveals certain research gaps yet to be addressed.
% which may hinder the current research from applying to analyzing human resource grouping based on event logs.
% This topic aligns with our focus of utilizing process data to support the analysis and improvement of resource grouping based on event log data. 
% The research presented in this paper focuses on the topic since our review of the literature of organizational model mining reveals major research gaps yet to be addressed. 
% 
%The evaluations conducted in the existing research rely on technique- or context-specific solutions, e.g., assessing the quality of the discovered resource groups using the performance measures specific to cluster analysis~\cite{jin2007organizational}, or referring to specific knowledge about the ``official'' organizational structure~\cite{song2008towards}. 
%with a focus on identifying the underlying, existing structural patterns of resource groups in an organization as reflected by historical process execution (i.e., how resources may be linked with one another as groups).

\begin{figure}[t!!!]
    \centering
    \includegraphics[width=\linewidth]{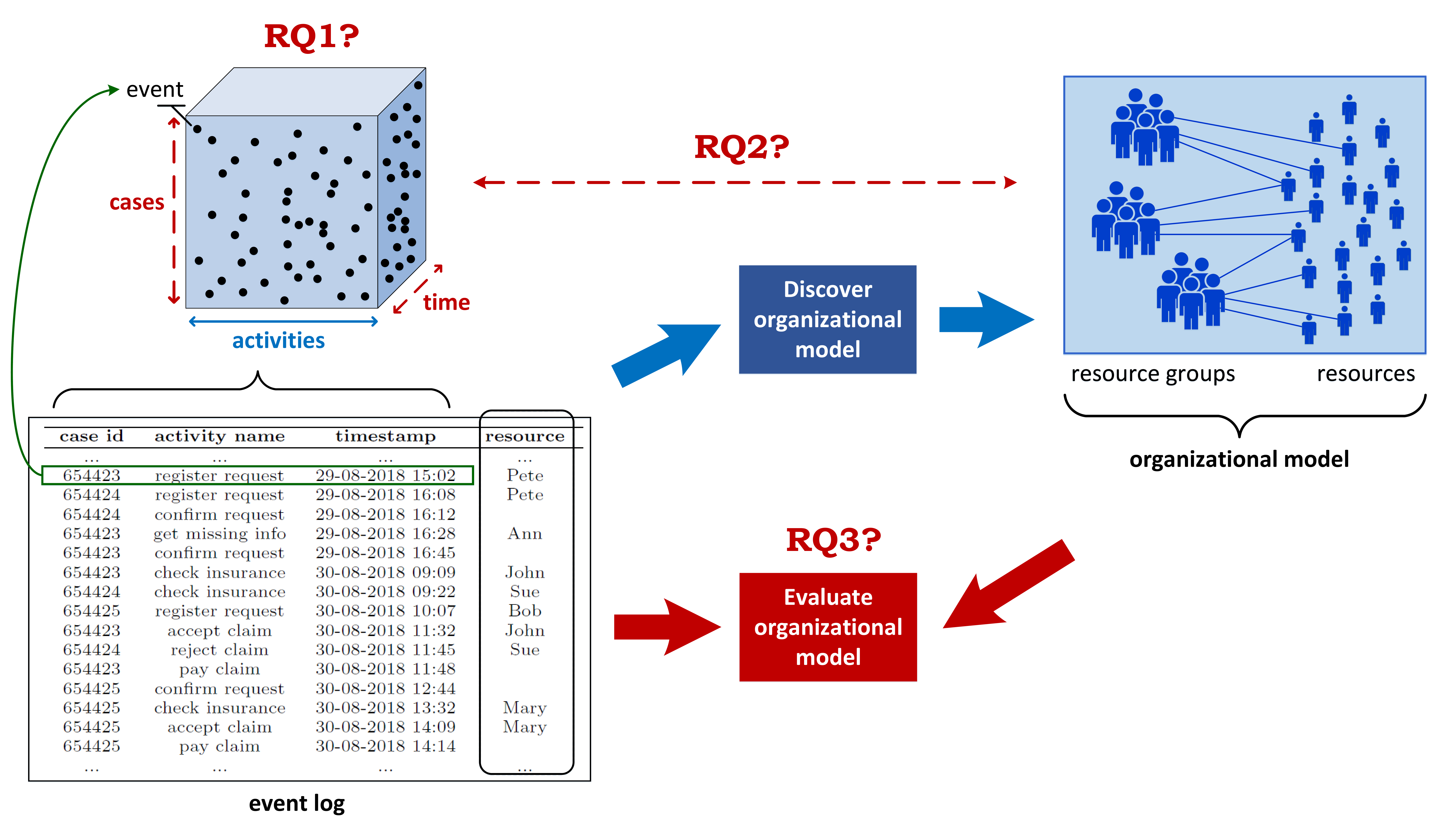}
    \caption{An illustration of the identified research questions in organizational model mining addressed in this paper.} %\todo{Decide between the two alternatives and remove the border fbox.}}
    \label{fig:intro}
\end{figure}

% Elaborate on the gaps/questions along with the figure components
\ifcomment %------------------- The whole paragraph updated --------------------%
Fig.~\ref{fig:intro} illustrates these open issues. 
For organizational model mining, typical input data from event logs often include information about the activities and human resources (in the form of resource identifiers), i.e., which resource carried out which activity in process execution. 
Additionally, an event log also records information relevant to the case and time aspects of process execution (case identifiers and possibly other case-level attributes, and timestamps). 
%In this regard, an event in an input log can be related to three basic dimensions of the process from a resource perspective: activity, case, and time.
In this regard, an event in an input log can be viewed as a data point carrying information on the three basic dimensions of the corresponding process from a resource perspective, i.e., activities, cases, and time.
% Gap #1
The existing studies proposed various methods for discovering organizational models from event logs, mainly based on how resources may differ in terms of their originated activities,
%~\cite{song2008towards,jin2007organizational,baumgrass2011deriving,zhao2012mining,burattin2013business,ye2018mining,ni2011mining,appice2018towards,yang2018finding,li2011a}
and how they are related due to the handover of work between activities.
%~\cite{zhao2012mining,burattin2013business,hanachi2012performative,sellami2012an,bouzguenda2015an}
On the other hand, the case and time dimensions are rarely concerned, thus limiting the possibilities of discovering resource groupings which follow patterns relevant to different cases, e.g., specialists group dedicated to specific sort of orders, or different time periods, e.g., employees with the same role but working different shifts.
% Gap #2: does it happen as a consequence of Gap #1, or is it just another separate issue?
Moreover, the discovered organizational models in the state-of-the-art papers often describe merely the clustering of resources, but lack in the connections from resource groups to the process underlying the event log data. Such a lack may pose a challenge for subsequent analysis of the behavior and performance of resource groups in process execution.
% Gap #3
Last but not least, in terms of evaluating the quality of the discovery results, the state-of-the-art rely on either extra domain knowledge or measures specific to the adopted techniques. The general idea shared by process mining research that evaluates a discovered model against the input event log remains to be explored in organizational model mining.
In light of these identified issues, we formulated the following research questions (RQ):
\textit{
\begin{enumerate}
    \item [RQ1.] How to incorporate event information on multiple dimensions into the discovery of organizational models from event logs?
    \item [RQ2.] How to connect resource groups in an organizational model with the business process underlying the input event log?
    \item [RQ3.] How to evaluate a discovered organizational model against the input log?
\end{enumerate}
}
\fi %------------------------------------------------------------------------------------%

Fig.~\ref{fig:intro} illustrates these open issues.  
%an event log is expected to carry resource information associated with an event, e.g., identifier of a resource who carried out a certain activity at some point of time during the execution of a particular case. 
Each event in an event log can be viewed as a data point carrying information on the three key dimensions of process execution, i.e., activities, cases, and time. 
% Gap #1
However, various methods proposed in the existing studies for discovering organizational models from event logs, are mainly based on how resources are similar to or differ from each other in terms of their originated activities, 
%~\cite{song2008towards,jin2007organizational,baumgrass2011deriving,zhao2012mining,burattin2013business,ye2018mining,ni2011mining,appice2018towards,yang2018finding,li2011a}
and how resources are related in terms of the handover of work between activities. 
%~\cite{zhao2012mining,burattin2013business,hanachi2012performative,sellami2012an,bouzguenda2015an}
On the other hand, the case and time dimensions are rarely considered, thus limiting the possibilities of discovering resource groupings which follow patterns relevant to different cases (e.g., specialists group dedicated to specific sort of orders) or different time periods (e.g., employees with the same role but working different shifts). 
% Gap #2: does it happen as a consequence of Gap #1, or is it just another separate issue?
Moreover, the discovered organizational models in the literature often describe merely the clustering of resources, but lack the connections between the identified resource groups and the process execution dynamics recorded in the event log data. Consequently, the missing connections pose a challenge for subsequent analysis of the behavior and performance of resource groups in the context of process execution. 
% Gap #3
Last but not least, in terms of evaluating the quality of the discovery results, the state-of-the-art relies on either domain knowledge or measures specific to the adopted techniques.
However, assessing the quality of a discovered organizational model by comparing it against the input event log has never been addressed. 
%For process mining research, it has been well-acknowledged that in terms of assessing a model discovered from an event log, one should first consider comparing the model against the source event log by evaluating their conformance. However, such an idea of evaluation remains to be explored in the literature on organizational model mining.
%\rev{The} idea of evaluating a discovered model against the input event log remains to be explored in organizational model mining, whereas in process mining research, it has been well-acknowledged that the assessment of models discovered from event logs should consider comparing models with logs (e.g., fitness~\cite{vanderaalst2016process}) as the first priority.
%The general idea shared by process mining research that evaluates a discovered model against the input event log remains to be explored in organizational model mining. 
In light of these identified issues, we formulate three research questions (RQ):
\textit{
\begin{enumerate}
    \item [RQ1.] How to incorporate event information on multiple dimensions into the discovery of organizational models from an input event log?
    \item [RQ2.] How to connect resource groups in an organizational model with the process execution information captured by the input event log?
    \item [RQ3.] How to evaluate a discovered organizational model against the input log?
\end{enumerate}
}

In this paper, we report our contributions that aim at addressing these research questions. 
%which addresses these research questions. 
We propose a novel framework built upon a richer definition of organizational models incorporating multiple dimensions of a business process and linking resource groups with process execution (addressing RQ1 and RQ2). Therefore, more in-depth analyses of resource groupings coupled with a process context may be enabled.
Through introducing the notions and measures for conformance checking, the proposed framework allows the rigorous evaluation of organizational models with reference to event log data, without being bound by how the models are constructed (addressing RQ3). 
We present an approach underpinned by the framework for discovering organizational models, 
%for discovering organizational models from a given event log, 
and conduct a series of experiments on real-life event logs to evaluate the proposed framework and approach in terms of both the discovery and conformance checking of organizational models. 
Open-source software was developed that implements the proposals in the paper.
We expect that the establishment of our framework will empower the use of process mining to support analyzing and improving resource grouping for organizational management. 
%and also unlock the potentials of process mining for future research with regard to human resources relevant to processes. 

%As an example, we demonstrate an approach under the framework for discovering organizational models from a given event log. Furthermore, 

% Roadmap for the paper
The remainder of the paper is structured as follows.
Section~\ref{sec:related} provides more detail about the state-of-the-art process mining research on the topic of organizational model mining.
Section~\ref{sec:framework} elaborates on the novel framework proposed in this paper. 
Subsequently, Section~\ref{sec:approach} presents an approach that realizes the discovery of organizational models discussed in the framework.
Section~\ref{sec:implementation} provides information about the related software implementation of our work.
Section~\ref{sec:evaluation} reports the evaluation of our work through experiments and discusses the findings.
Finally, Section~\ref{sec:conclusion} summarizes the paper and presents an outlook for future research topics that can be built upon the current work.

%while lacks in a general and rigorous way of evaluating organizational models against the source process-related data (as the conformance checking for process model discovery~\cite{rozinat2008,adriansyah2013,Garcia-Banuelos2018}). 
%This implies yet another gap that may hinder process mining research from applying to the study of resource grouping in business process context. 

%Consequently, it remains unclear how to interpret the derived results; also, such a missing link between resource groups and processes may prevent one from gaining insights on further revisions through incorporating process knowledge. 

%Such a lack leaves the conducted evaluations ill-posed: context-specific prior knowledge could indeed be different from the \textit{de facto} behavior recorded in event logs as organizational structure may have evolved while processes do; and technique-specific evaluation could be biased as it depends on the nature of specific technique adopted. Comparing organizational models with event logs as the reference is needed for effective evaluation of any mining result and thus ensuring the value of subsequent analysis based on it.

% State-of-the-art organizational model mining: what are the gaps
% Nevertheless, certain gaps exist which hinder the current research from being applied to tackle the aforementioned challenge. 

\section{Related Work}
\label{sec:related}

%A very brief overview of organizational mining and hence introduce organizational model mining.

Process mining research that aims at deriving human-resource-related insights from event log data is known as organizational mining~\cite{song2008towards}. Four topics of interest can be identified, respectively:
(1) {organizational model mining}, which aims at discovering knowledge related to organizational structures around human resources (e.g.,~\cite{song2008towards,yang2018finding}); 
(2) {social network mining}, which aims at extracting and analyzing the relationships among resources (e.g.,~\cite{vanderaalst2004mining,vanderaalst2005discovering,liu2013accelerating}); 
(3) {rule mining}, which aims at extracting inherent rules that decide the use of human resources in process execution, e.g., rules for the assignment of tasks to resources~\cite{ly2005mining,rinderlema2007,huang2011mining,schonig2016a}, and rules for the composition of resource teams with various expertise~\cite{schonig2016b}; and 
(4) {behavioral profile mining}, which aims at extracting and analyzing different aspects of individual resource behavior as they participate in process execution (e.g.,~\cite{nakatumba2009analyzing,pika2017mining,huang2012resource,suriadi2017discovering}).

The focus of this paper is on organizational model mining. To review and evaluate the related work, we establish certain criteria from three perspectives based on the research questions raised in Section~\ref{sec:intro}. %the aforementioned research questions.
% 1. Does the paper consider one or more aspects of process execution for discovering organizational models?
The first perspective concerns the different aspects of process execution considered for organizational model discovery. A typical event log suitable for conducting organizational model mining often records the minimum information, including activity labels, resource identifiers, case identifiers, and timestamps. Using the information, the participation of human resources in process execution can be analyzed from several aspects in order to discover organizational models, i.e., how resources carry out activities (activity), how they are involved in different cases (case), how they work in various time periods (time), and how they interact during process execution (resource interaction). 
% 2. Does a discovered organizational model in the paper include linkage with the business process?
The second perspective concerns whether or not a discovered model specifies linkage between resource groups and the process execution dynamics captured by the input event log. 
Considering the three key dimensions of process execution, a resource group in a discovered model may carry out certain activities, get involved in certain cases, and work during certain time periods. 
%can be related to some aspects of the process, for instance, specific activities or time periods.
% 3. What strategy is employed in the paper for evaluating the discovery method?
The third perspective concerns how the evaluation of results is carried out. 
For this perspective, we take into account several common practices, including evaluating a discovered model against the input event log, evaluating a discovered model against relevant prior knowledge, and measuring the effectiveness of the applied technique.

Based on these criteria, Table~\ref{tab:related_work_omm} lists the state-of-the-art research on discovering organizational models from event logs.
\newline

\begin{table}[hb]
\caption{The state-of-the-art on discovering organizational models from event logs.}
\label{tab:related_work_omm}
\centering
\begin{footnotesize}
    \resizebox{\linewidth}{!} {
    \begin{tabular}{|c|cccc|c|cccc|}
        \hline
        \multirow{3}{*}{Ref.} &  \multicolumn{4}{c|}{Aspects of process execution} & \multirow{3}{*}{Links to process} & \multicolumn{4}{c|}{Evaluation strategy} \\
        \cline{2-5} \cline{7-10}
        & \multirow{2}{*}{activity} & \multirow{2}{*}{case} & \multirow{2}{*}{time} & resource & & prior & event & technique & feasibility \\
        & & & & interaction & & knowledge & log & effectiveness & validation \\
        \hline
        
        \cite{song2008towards} & $\surd$ & $\surd$ & & & $\surd$ & $\surd$ & & & \\
        \hline
        \cite{jin2007organizational} & $\surd$ & & & & & $\surd$ & & & \\
        \hline
        \cite{baumgrass2011deriving} & $\surd$ & & & & $\surd$ & & & & $\surd$ \\
        \hline
        \cite{zhao2012mining} & $\surd$ & & & $\surd$ & $\surd$ & & & & $\surd$ \\
        \hline
        \cite{burattin2013business} & $\surd$ & & & $\surd$ & $\surd$ & $\surd$ & & & \\
        \hline
        \cite{ye2018mining} & $\surd$ & & & & & & & $\surd$ & \\
        \hline
        \cite{ni2011mining} & $\surd$ & & & & & & & & $\surd$ \\
        \hline
        \cite{appice2018towards} & $\surd$ & & & & & & & $\surd$ & \\
        \hline
        \cite{yang2018finding} & $\surd$ & & & & & $\surd$ & & & \\
        \hline
        \cite{li2011a} & $\surd$ & & & & & $\surd$ & & & \\
        \hline
        \cite{hanachi2012performative} & & & & $\surd$ & & & & & $\surd$ \\
        \hline
        \cite{sellami2012an} & & & & $\surd$ & & & & & $\surd$ \\
        \hline
        \cite{bouzguenda2015an} & & & & $\surd$ & & & & & $\surd$ \\
        \hline
    \end{tabular}
    }
\end{footnotesize}
\end{table}

\noindent \textit {
1. Does the paper consider one or more aspects of process execution for discovering organizational models?
}

Our review of the literature shows that the majority of the proposed methods utilize information about resources carrying out different activities for discovering organizational models. The idea is motivated by the observation that common resource grouping schemes (e.g., by roles, functional units) often lead to different groups of employees being in charge of executing some particular activities in a process.
Some research also exploits the information on the interaction among resources (e.g., handover of work between the execution of activities) for discovering organizational models, especially those focusing on describing the reporting relationships~\cite{hanachi2012performative,sellami2012an,bouzguenda2015an}.
On the other hand, information related to cases (i.e., process instances) and time is rarely considered. Only reference~\cite{song2008towards} attempts to exploit case information in event logs for discovering organizational models reflecting project teams or task force, which consists of employees with various specialties and assembled for collective tasks. 
For most of the current work, the discovery of organizational models takes into account merely the activity information in event logs.
As a consequence, these models may not be able to capture resource groupings that follow certain patterns on the case and time dimensions. 
The various aspects that can be utilized for characterizing resource participation in process execution are yet to be explored. %fully 
\newline

\noindent \textit{
2. Does a discovered organizational model in the paper specify links to the execution of the business process captured by the input event log?
}

All organizational models derived from event logs in the referenced papers describe the grouping structures around or the reporting relationships among human resources. 
Yet only a few of them~\cite{song2008towards,baumgrass2011deriving,zhao2012mining,burattin2013business} attach a discovered model to the business process captured by the event log data, which is done by linking the resource groups with process activities for characterizing the responsibilities or permissions of the groups.
In order to analyze the behavior and performance of resource groups in a discovered organizational model, it is necessary to situate the groups in the context of process execution. 
%By doing so, it may also contribute to further utilizing organizational models for supporting better planning of human resource deployment~\cite{pika2017mining}. 
However, the issue remains open in the current organizational model mining research.
\newline

\noindent \textit{
3. What strategy is employed in the paper for evaluating the results?
}

%In terms of 
With respect to the evaluation part, there exist three typical strategies in the current research. The first one is to compare the discovery results with domain knowledge, such as the official organizational structure~\cite{song2008towards,jin2007organizational,burattin2013business,yang2018finding,li2011a}. It relies on the availability of such knowledge, and may also be flawed as there is no guarantee whether or not the reality (event log data) has deviated from the referenced domain knowledge.
The second solution is to assess the effectiveness of the techniques applied for model discovery, for instance, the quality of the detected resource communities measured by modularity~\cite{ye2018mining,appice2018towards}. Such a solution may be biased since the assessing criteria are often dependent on the techniques selected.
The last one considers evaluation without assessing the quality of the discovered models, but instead only validates the feasibility of the proposed methods through implementation and experimentation on synthetic or real-life event logs ~\cite{baumgrass2011deriving,zhao2012mining,ni2011mining,hanachi2012performative,sellami2012an,bouzguenda2015an}. 
%This may raise questions around the utility of the methods in different scenarios.
%
So far, none of the existing studies on organizational model mining has explicitly considered using the input event logs as references for evaluating the discovered models.
\newline
%despite the fact that such an idea has been well-acknowledged by process mining research.

In this paper, we target these three open issues in the current organizational model mining research, and address them through proposing a novel framework consisted of a new definition of organizational models as well as the notions of conformance checking for organizational models. The next section elaborates on our proposals.

\section{Framework} 
\label{sec:framework}

A business process is formed by a set of \textit{activities} to achieve a business goal (e.g., to handle an insurance claim). It captures possible alternative ways of performing the activities to achieve a business goal 
(e.g., a sequence of activities required to handle an insurance claim for VIP customers is different from that for normal customers). 
%cater for different context of its business goal 
An instance of executing a process is known as a \textit{case}, and a collection of cases sharing certain common characteristics may exemplify a type of case (or a \textit{case type}), e.g., cases of handling insurance claims of VIP customers \textit{vs.}\ cases for normal customers. For process execution, many business activities are performed by \textit{resources} that may have different roles and positions in an organization and form \textit{resource groups} that may reflect various organizational entities, e.g.,\ an organizational department or a project team.   
%belong to different organizational units (in form of \emph{resource groups}). 

In this section, we introduce the notion of a richer \textit{organizational model} linking resources via resource groups to activities, case types, and time periods associated with process execution. Building upon this notion, we present a framework which supports discovering such organizational models from event logs, and evaluating them by conformance checking with reference to the actual behavior of resources as recorded in the logs. %grouping. 

\subsection{Event Log}
\label{sec:framework/event_log}

% \todo{Most preliminaries are adopted directly from Wil's document on OrgMining 2.0. Modify them when necessary.}
% \todo{Introduction of event logs as a common way of recording process event data and the background knowledge.}

An event log consists of a set of events, and each event is associated with a range of event attributes capturing the information about carrying out a process activity. 
For example, Table~\ref{tab:example_event_log} lists a small fragment of some event log. 
Each row corresponds to an event, and each column records, for each event, the value of an event attribute of which the attribute name is shown as the column header.

\begin{table}[htb]
\caption{A fragment of some event log.}%: each line corresponds to an event and each row to an event attribute} 
\label{tab:example_event_log}
\centering
\begin{scriptsize}
    \begin{tabular}{cccccc}
        \hline
        \bf case id &  \bf activity name & \bf timestamp & \bf resource & \bf customer type \\
        \hline
				... & ... & ... & ... & ... \\
        654423 & register request & 29-08-2018 15:02 & Pete & normal \\
        654424 & register request & 29-08-2018 16:08 & Pete & normal \\
        654424 & confirm request & 29-08-2018 16:12 &  & normal \\
        654423 & get missing info & 29-08-2018 16:28 & Ann & normal \\
        654423 & confirm request & 29-08-2018 16:45 &  & normal \\
        654423 & check insurance & 30-08-2018 09:09 & John & normal \\
        654424 & check insurance & 30-08-2018 09:22 & Sue & normal \\
        654425 & register request & 30-08-2018 10:07 & Bob & VIP \\
        654423 & accept claim & 30-08-2018 11:32 & John & normal \\
        654424 & reject claim & 30-08-2018 11:45 & Sue & normal  \\
        654423 & pay claim & 30-08-2018 11:48 &  & normal  \\
        654425 & confirm request & 30-08-2018 12:44 &  & VIP \\
        654425 & check insurance & 30-08-2018 13:32 & Mary & VIP \\
        654425 & accept claim & 30-08-2018 14:09 & Mary & VIP \\
        654425 & pay claim & 30-08-2018 14:14 &  & VIP \\
				... & ... & ... & ... & ... \\
        \hline
    \end{tabular}
\end{scriptsize}
\end{table}

We first define a general data structure of an event log (see Def.~\ref{def:event_log}). 
An event log ($\mi{EL}$) contains the information of a set of uniquely identifiable events ($E$), a set of event attribute names ($\mi{Att}$), and the corresponding event attribute values carried by each event (as specified by function~$\pi$). 
It is possible that an event does not carry any value for a given event attribute (e.g., in Table~\ref{tab:example_event_log} several events have no resource information). 
Hence, function~$\pi$ maps the set of events~$E$ to a partial function~$\mi{Att}\not\rightarrow\univ_{\mi{Val}}$ and, as such, given any event ($e\in E$) only the attributes that carry a value ($\mi{dom}(\pi(e))$) are  presented for the event. 

\begin{definition}[Event Log]\label{def:event_log}
    Let $\mathcal{E}$ be the universe of event identifiers, $\univ_{\mi{Att}}$ be the universe of possible attribute names, and $\univ_{\mi{Val}}$ the universe of possible attribute values.
    $\mi{EL}=(E,\mi{Att},\pi)$ with $E \subseteq {\cal E}$, $\mi{Att} \subseteq \univ_{\mi{Att}}$, and $\pi \in E \rightarrow (\mi{Att} \not\rightarrow \univ_{\mi{Val}})$ is an event log.
    Event $e \in E$ has attributes $\mi{dom}(\pi(e))$. For $x \in \mi{dom}(\pi(e))$, $\pi_x(e) = \pi(e)(x)$ is the value of attribute $x$ for event $e$.
    \hfill\qed  
\end{definition}

Next, we elaborate on the definition of event attributes that are needed for storing essential information about process execution (see Def.~\ref{def:event_attributes}). 
An event log often records many cases (i.e., process execution instances). 
Each case is uniquely identifiable and consists of a sequence of events corresponding to the execution of activities at some specific time. 
As the minimum requirement for event logs, events have three mandatory attributes~\cite{vanderaalst2004workflowmining}: \textit{case identifier} (\textit{case} for short), \textit{activity name} (\textit{act} for short), and \textit{timestamp} (\textit{time} for short). 
Optionally, each event may have a corresponding resource executing it, and hence \textit{resource} (\textit{res} for short) is another important attribute. 
In addition to these four most common ones, an event log may record other event attributes (such as \textit{customer type}, \textit{cost}, \textit{department}, etc.), which are usually non-mandatory and vary between event logs. The event log fragment shown in Table~\ref{tab:example_event_log} is a good example of an event log that contains the above event attributes. 

% [TODO] Mention Eres for the set of events that have associated resources information ... given the focus of this paper we are interested in these events

\begin{definition}[Event Attributes]\label{def:event_attributes}
    Let $\mathcal{C} \subseteq \univ_{\mi{Val}}$, $\mathcal{A} \subseteq \univ_{\mi{Val}}$,  $\mathcal{T} \subseteq \univ_{\mi{Val}}$ and $\mathcal{R} \subseteq \univ_{\mi{Val}}$ denote the universe of case identifiers, the universe of activity names, the universe of timestamps, and the universe of resource identifiers, respectively. Any event log $\mi{EL}=(E,\mi{Att},\pi)$ has four special attributes: $\{\mi{case},\mi{act},\mi{time},\mi{res} \} \subseteq \mi{Att}$ such that for any $e \in E$:
    \begin{itemize}
    	\vspace*{-.2\baselineskip}
    	\setlength\itemsep{-0.2em}
        \item $\{\mi{case},\mi{act},\mi{time}\} \subseteq \mi{dom}(\pi(e))$, i.e., these attributes are mandatory,
        \item $\pi_{\mi{case}}(e) \in \mathcal{C}$ is the case to which $e$ belongs,
        \item $\pi_{\mi{act}}(e) \in \mathcal{A}$ is the activity $e$ refers to,
        \item $\pi_{\mi{time}}(e) \in \mathcal{T}$ is the time at which $e$ occurred, and
        \item $\pi_{\mi{res}}(e) \in \mathcal{R}$ is the resource that executed $e$ if $\mi{res} \in \mi{dom}(\pi(e))$ (this attribute is optional).
    	\vspace*{-.2\baselineskip}
    \end{itemize}
	%\hspace*{1em}
	Hence, $E_{\mi{res}} = \{ e \in E \mid \mi{res} \in \mi{dom}(\pi(e))\}$ are the events executed by a resource and $E_{\mi{nres}} = \{ e \in E \mid \mi{res} \not\in \mi{dom}(\pi(e))\}$ are all other events.\hfill\qed
\end{definition}

\subsection{Execution Mode} 
\label{sec:framework/exec_mode}

%We are interested in recognizing groups of resources having similar characteristics.  
In the context of business processes, the behavior of resources or resource groups is often closely associated with certain process features.  
For example, resources belonging to a certain resource group may perform (only) a specific type of activity or case. 
Consider the event log fragment in Table~\ref{tab:example_event_log}. It records the execution of an insurance claim process. 
By grouping the events using attribute ``activity name'', we can tell that Pete and Bob could only ``register (a claim) request'', and John, Sue, and Mary could ``check insurance'' and then decide to ``accept (or reject a) claim''. 
By grouping the events using attribute ``customer type'', we can observe that Bob and Mary only handled a claim made by a VIP customer, while the other resources are responsible for dealing with claims of normal customers. 
Based on these observations, we may infer potential grouping of resources relevant to activity name or customer type. 
To reach such findings in this example, an important step is the classification of the individual events in an event log into groups of events characterized by relevant event attribute(s).  

Let us first introduce the concepts of \emph{case types}, \emph{activity types}, and \emph{time types} (see Def.~\ref{def:cat_types}). 
These are defined with respect to the universe of possible attribute values for each of the three mandatory event attributes (\emph{case}, \emph{act}, and \emph{time}) of an event log, respectively. 
Case types correspond to the classification of cases and are often informed by relevant case-level event attribute(s) (e.g., customer type, cost, urgency).  
In the above example, there are two types of cases characterized by different types of customers -- claims by VIP customers and claims by normal customers.  
From the data viewpoint, a collection of \emph{case types} is related to a way of partitioning the universe of case identifiers (uniquely representing individual cases).  
It means that each case type corresponds to a cluster of cases, while each case has only one case type. 
The same principle applies to the definition of \emph{activity types} and \textit{time types}. 
Examples of activity types are registration, approval, payment, etc., and examples of time types are weekdays, weekend, morning, afternoon, etc.  

%For example, we may create two time types corresponded to either weekdays or weekends, and any timestamp of event data will be uniquely designated as being one of the time types (and not the other). Def.~\ref{def:cat_types} formalizes the notion as independent of particular ways to construct such partitions.

\begin{definition}[Case Types, Activity Types, and Time Types]\label{def:cat_types}
    Let $\mathcal{CT}$, $\mathcal{AT}$, and $\mathcal{TT}$ denote the sets of names of case types, activity types, and time types, respectively.
    For any $\mathcal{CT}$, $\mathcal{AT}$, or $\mathcal{TT}$, there exists an injective function\footnote{A function $f: X \rightarrow Y$ is said to be injective if and only if every element of $Y$ is the image of at most one element of $X$.}:
    \begin{itemize}
        \vspace*{-.2\baselineskip}
        \setlength\itemsep{-0.2em}
        \item $\varphi_\mi{case}: \mathcal{CT} \rightarrow \powerset(\mathcal{C})$
        \footnote{Given a set $S$, $\powerset(S)$ is the powerset of $S$, i.e., the set of all subsets of $S$. $X\in\powerset(S)$ if and only if $X\subseteq S$.} 
        satisfying \ 
        $\bigcup{\mi{rng}(\varphi_\mi{case})} = \mathcal{C}, \forall_{\mi{ct}_1,\mi{ct}_2 \in \mathcal{CT}}( \varphi_\mi{case}{(\mi{ct}_1)} \cap \varphi_\mi{case}{(\mi{ct}_2)} = \emptyset \lor \mi{ct}_1 = \mi{ct}_2 )$;
        \item $\varphi_\mi{act}: \mathcal{AT} \rightarrow \powerset(\mathcal{A})$ satisfying \ 
        $\bigcup{\mi{rng}(\varphi_\mi{act})} = \mathcal{A}, \forall_{\mi{at}_1,\mi{at}_2 \in \mathcal{AT}}( \varphi_\mi{act}{(\mi{at}_1)} \cap \varphi_\mi{act}{(\mi{at}_2)} = \emptyset \lor \mi{at}_1 = \mi{at}_2 )$;
        \item $\varphi_\mi{time}: \mathcal{TT} \rightarrow \powerset(\mathcal{T})$ satisfying \ 
        $\bigcup{\mi{rng}(\varphi_\mi{time})} = \mathcal{T}, \forall_{\mi{tt}_1,\mi{tt}_2 \in \mathcal{TT}}( \varphi_\mi{time}{(\mi{tt}_1)} \cap \varphi_\mi{time}{(\mi{tt}_2)} = \emptyset \lor \mi{tt}_1 = \mi{tt}_2 )$.
        \vspace*{-.2\baselineskip}
    \end{itemize}
    i.e., any \ $\mathcal{CT}$, $\mathcal{AT}$, or $\mathcal{TT}$ is related to a way of partitioning the corresponding universe.
    %These collections partition the corresponding universes, i.e., 
	%	$\mathcal{CT}\subseteq \powerset(\mathcal{C})$\footnote{Given a set $S$, $\powerset(S)$ is the powerset of $S$, i.e., the set of all subsets of $S$. $X\in\powerset(S)$ if and only if $X\subseteq S$.} such that $\bigcup \mathcal{CT} = \mathcal{C}$, $\forall_{C_1,C_2 \in \mathcal{CT}}\ C_1 \cap C_2 = \emptyset \lor C_1 = C_2$,
    %$\mathcal{AT}\subseteq \powerset(\mathcal{A})$ such that $\bigcup \mathcal{AT} = \mathcal{A}$, $\forall_{A_1,A_2 \in \mathcal{AT}}\ A_1 \cap A_2 = \emptyset \lor A_1 = A_2$,
    %$\mathcal{TT}\subseteq \powerset(\mathcal{T})$ such that $\bigcup \mathcal{TT} = \mathcal{T}$ and $\forall_{T_1,T_2 \in \mathcal{TT}}\ T_1 \cap T_2 = \emptyset \lor T_1 = T_2$.
    \hfill\qed
\end{definition}                        
%
%\begin{footnotesize}
%[Note] Given any set $S$, $\powerset(S)$ is the powerset of $S$, i.e. all subsets of $S$. $X\in\powerset(S)$ if and only if $X\subseteq S$.\\ \vspace*{-.5\baselineskip}
%\end{footnotesize}

We then define the notion of \emph{execution mode} (see Def.~\ref{def:exec_mode}). 
An execution mode refers to the possible execution of events characterized by the combination of a case type, an activity type, and a time type. 
For example, it can be the execution of a ``registration activity'' (activity type) in a ``case related to VIP customers'' (case type) on a ``weekday morning'' (time type). Execution modes partition the universe of possible events. Each execution mode corresponds to a collection of events, and each event has precisely one execution mode. 

\begin{definition}[Execution Mode]\label{def:exec_mode}
$(\mi{ct},\mi{at},\mi{tt}) \in \mathcal{CT} \times \mathcal{AT} \times \mathcal{TT}$ is an execution mode.
    It refers to the possible execution of event $e$ satisfying
    $\pi_{\mi{case}}(e) \in \varphi_\mi{case}{(\mi{ct})}$,
    $\pi_{\mi{act}}(e) \in \varphi_\mi{act}{(\mi{at})}$, and
    $\pi_{\mi{time}}(e) \in \varphi_\mi{time}{(\mi{tt})}$. 
Given any event log $\mi{EL}=(E,\mi{Att},\pi)$, 
$[E]_{(\mi{ct},\mi{at},\mi{tt})} = \{ e \in E \mid \pi_\mi{case}(e) \in \varphi_\mi{case}{(\mi{ct})} \land \pi_\mi{act}(e) \in \varphi_\mi{act}{(\mi{at})} \land \pi_\mi{time}(e) \in \varphi_\mi{time}{(\mi{tt})}\}$ is the set of events having execution mode $(\mi{ct},\mi{at},\mi{tt})$ in $\mi{EL}$.\hfill\qed 
%$[E_\mi{res}]_{(\mi{ct},\mi{at},\mi{tt})} = \{ e \in E_\mi{res} \mid \pi_\mi{case}(e) \in \mi{ct} \land \pi_\mi{act}(e) \in \mi{at} \land  \pi_\mi{time}(e) \in \mi{tt}\}$  
\end{definition}

\begin{figure}[t!!!]
\centering
    \subfloat[An event is seen as a dot in the three dimensions of cases, activities, and time, and may have a resource executing it.]
		{\includegraphics[width=3in]{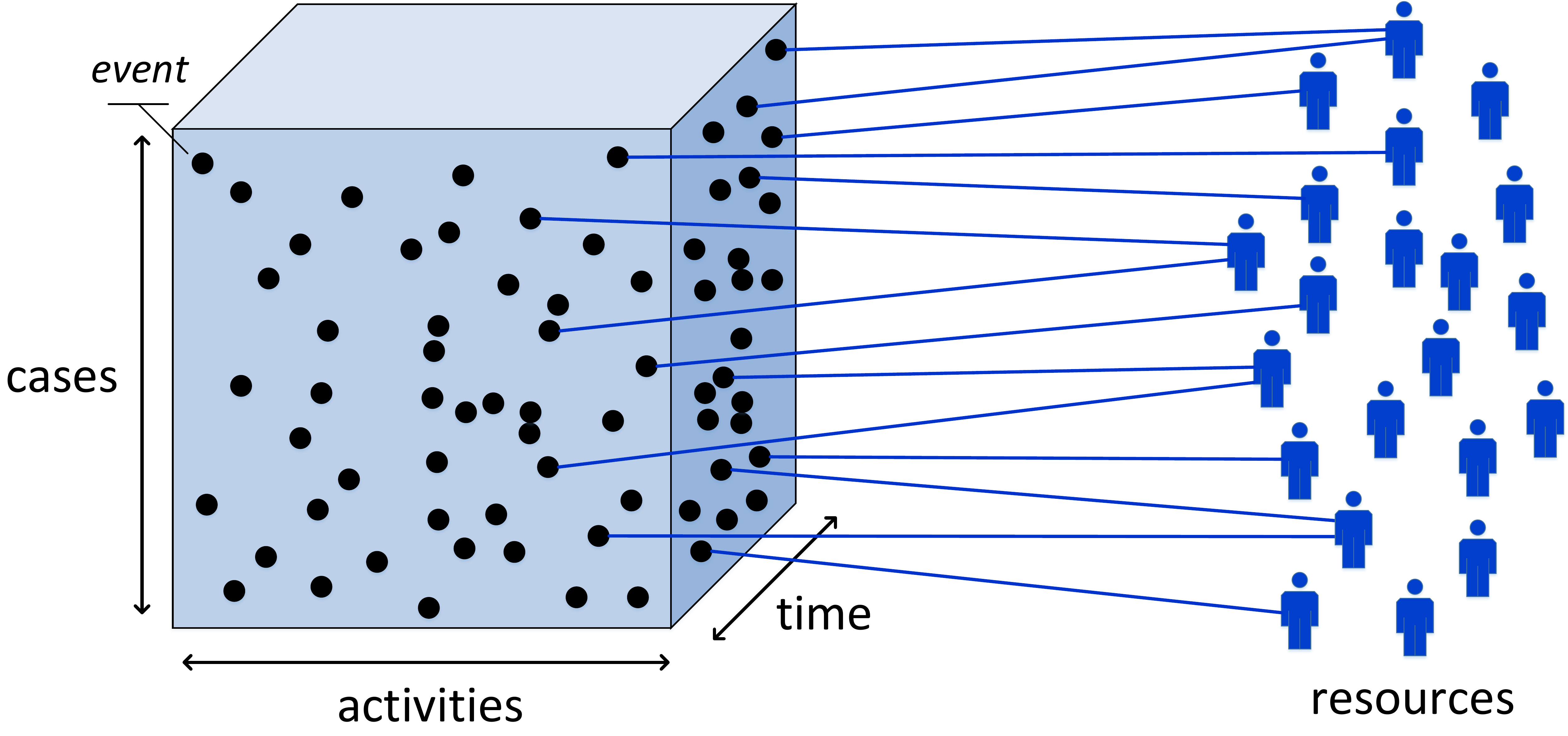} \label{fig:log-view/event-3d}}\\
    \subfloat[An execution mode is seen as an event cube in the three dimensions of case types, activity types, and time types, and may have resources executing the events in the cube.]
		{\includegraphics[width=3.02in]{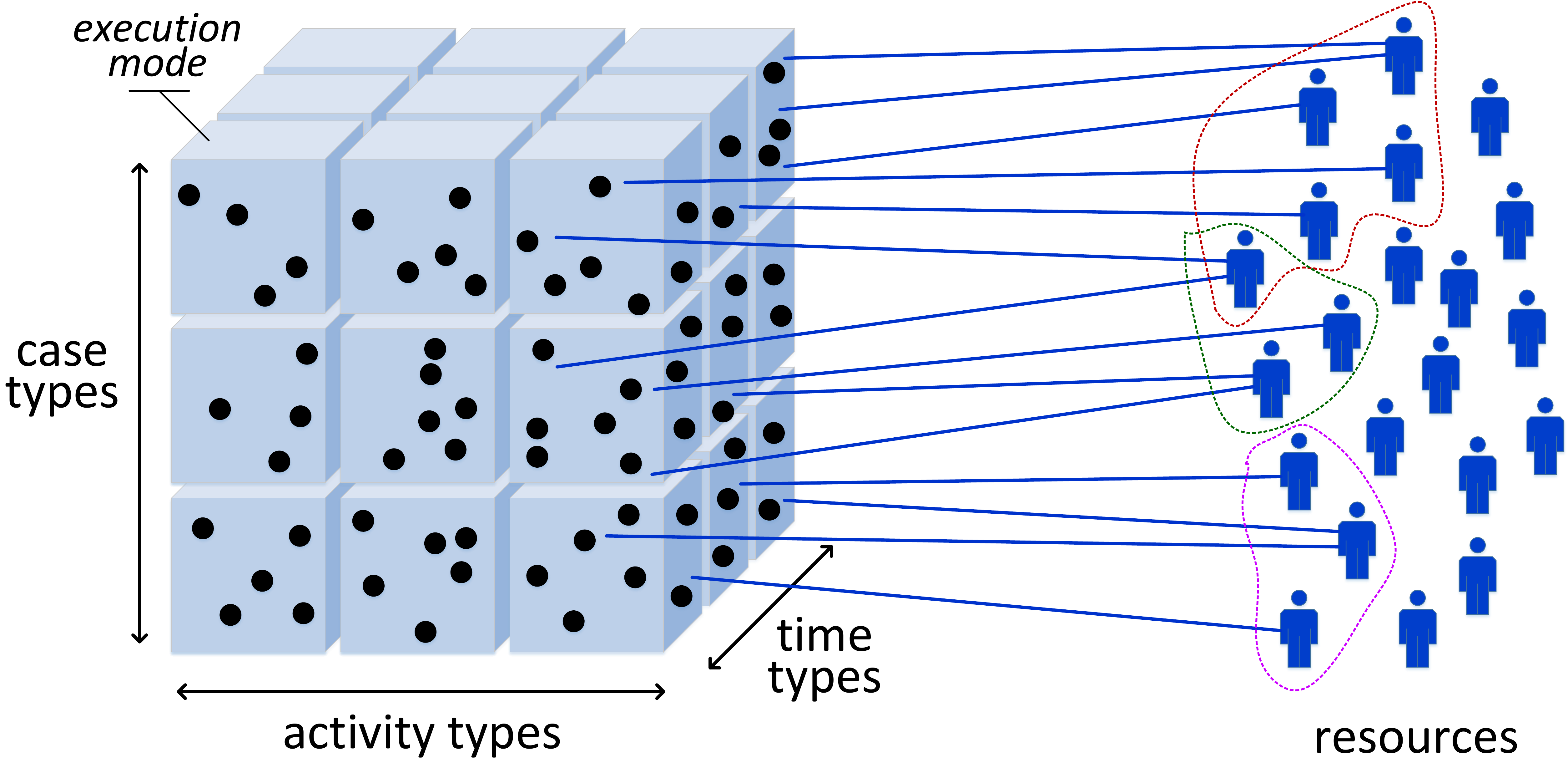} \label{fig:log-view/event_cube}}
\caption{Graphical views of (a) events and (b) execution modes and, optionally, the associated resource information recorded in an event log.}% -- in a three dimensional space and optionally linked to the corresponding resources}% executing some of them}
%positioned along three dimensions of cases, activities and time, and optionally linked to a corresponding resource executing it; and (b)}
%\caption{Two graphical views: (a) events positioned along three dimensions of cases, activities and time, and optionally linked to a corresponding resource executing it; and (b)}
\label{fig:log-view}
\end{figure}

Fig.~\ref{fig:log-view} presents two graphical views of events to illustrate the above concepts. 
In~Fig.~\ref{fig:log-view/event-3d}, events are seen as dots in a three-dimensional space, which represents the three mandatory event attributes capturing the information of \emph{cases}, \emph{activities}, and \emph{time}. %, and an event log containing all the events can be seen as one entire cube (which we refer to as an event log cube). 
In~Fig.~\ref{fig:log-view/event_cube}, the values along each of the three dimensions are partitioned according to the classification by case types, activity types, and time types, respectively.  As a result, the collection of all possible events are partitioned into a set of execution modes that can be viewed as event cubes formed by the combination of a case type, an activity type, and a time type.

%TBC: A set of execution modes specifies a partitioning of the space, and each event can therefore be described by the combination of a case type, activity type and time type.} 

\subsection{Resource Log}
\label{sec:framework/res_log}

%Different attributes of events can be utilized to conduct analysis using various process mining techniques~\cite{vanderaalst2016process}. To do so we adopt an idea similar to that of a \textit{process cube view}~\cite{vanderaalst2013process} to help develop a structured view on event data.
The focus of research is to extract the organizational knowledge about resource grouping, given the information recorded in event logs. Hence, we are more interested in viewing event data from the \emph{resource perspective}. 
In an event log, an event may be executed by a resource, and each resource is related to one or more individual events performed by the resource (see the graphical view in Fig.~\ref{fig:log-view/event-3d}). Consider an event log consisting of a set of execution modes. 
On the one hand, an execution mode corresponds to a collection of events and thus may be associated with one or more resources that have performed the relevant events. 
On the other hand, each resource may have executed several events that belong to different execution modes and thus may be linked to these (multiple) execution modes (see Fig.~\ref{fig:log-view/event_cube}).  
%In event logs, an event may have the information about the corresponding resource executing the event, and thus each resource is linked to individual events (see a graphical view in Fig.~\ref{fig:log-view/event-3d}). As discussed before, the grouping of resources is often based on certain characteristics informed by execution modes. This means that the resources belonging to the same resource group would most likely perform the events having the same execution modes. 
%Hence, each resource is involved in performing one or several execution modes, and as such each resource can be linked to execution modes instead of events (see Fig.~\ref{fig:log-view/event_cube}). 
In addition, Fig.~\ref{fig:log-view/event_cube} also depicts some potential grouping of resources, assuming that resources form groups based on case types. 
%the %performing the cases of the same case type form a group. 

We introduce the term ``resource event'' to refer to an event executed by some resource and characterized by the corresponding execution mode. 
A so-called \emph{resource log} consists of a multiset of resource events (see Def.~\ref{def:res_log}), i.e., it may contain multiple resource events referring to the same resource and the same execution mode.
A resource log can be derived from an event log and a collection of pre-defined execution modes (via combinations of case types, activity types, and time types) (see Def.~\ref{def:derived_res_log}).  

\begin{definition}[Resource Log] \label{def:res_log}
    Let $\mathcal{R}$, $\mathcal{CT}$, $\mathcal{AT}$, and $\mathcal{TT}$ represent collections of resources, case types, activity types, and time types, respectively.
    $(r,\mi{ct},\mi{at},\mi{tt}) \in \mathcal{R} \times \mathcal{CT} \times \mathcal{AT} \times \mathcal{TT}$ is a resource event.
    A resource log $\mi{RL} \in \bag(\mathcal{R} \times \mathcal{CT} \times \mathcal{AT} \times \mathcal{TT})$\footnote{Given a set $S$, $\bag(S)$ is the set of all multisets over $S$.} 
    is a multiset of resource events.\hfill\qed
\end{definition}

\begin{definition}[Derived Resource Log] \label{def:derived_res_log}
    Let $\mi{EL}=(E,\mi{Att},\pi)$ be an event log and $\mi{CT}$, $\mi{AT}$, and $\mi{TT}$ be some defined collections of case types, activity types, and time types, respectively. 
    The resource log derived from $\mi{EL}$, $\mi{CT}$, $\mi{AT}$, and $\mi{TT}$ is %\\
    $\mi{RL}(\mi{EL},\mi{CT},\mi{AT},\mi{TT}) =
    [
    %(\mi{r}, \mi{ct}, \mi{at}, \mi{tt})
    (\pi_{\mi{res}}({e}), \mi{ct}, \mi{at}, \mi{tt})
    \mid
    %(\mi{r}, \mi{ct}, \mi{at}, \mi{tt}) \in \mathcal{R} \times \mi{CT} \times \mi{AT} \times \mi{TT} \land
    %\exists_{e \in [E_\mi{res}]_{(\mi{ct},\mi{at},\mi{tt})}} \pi_{\mi{res}}(e) = r
    e \in E_\mi{res} \land
    (\mi{ct},\mi{at},\mi{tt}) \in \mi{CT} \times \mi{AT} \times \mi{TT} \land
    e \in [E]_{(\mi{ct},\mi{at},\mi{tt})}
    ].$\footnote{$[\ldots \mid \ldots]$ generates a multiset where for every ``binding'' on the right-hand side, an element is added on the left-hand side. For example, $[x + 1 \mid x \in [1^2, 2^3, 5^4] \land x < 4] = [2^2, 3^3]$.}\hfill\qed
    %\footnote{$[\ldots \mid \ldots]$ builds a multiset which consists of all values that evaluate to true for the predicate on the right-hand. For example, $[x \mid x \in [1^2, 2^3, 5^4] \land x < 4] = [1^2, 2^3]$.}
\end{definition}

% ------------------- Definitions using a counter instead of a multiset notation ---------------
\ifcomment
\begin{definition}[Resource Log]
    Let $\mathcal{R}$, $\mathcal{CT}$, $\mathcal{AT}$ and $\mathcal{TT}$ represent collections of resources, case types, activity types, and time types, respectively.
    $(r,\mi{ct},\mi{at},\mi{tt}) \in \mathcal{R} \times \mathcal{CT} \times \mathcal{AT} \times \mathcal{TT}$ is a resource event.

    A resource log $\mi{RL}=(\mi{RE}, m)$ is a multiset of resource events, where $\mi{RE}\in\powerset(\mathcal{R} \times \mathcal{CT} \times \mathcal{AT} \times \mathcal{TT})$ is the set of (distinct) resource events and $m\in\mi{RE}\rightarrow\mathbb{N}^+$ specifies the number of occurrences of each resource event. 
% $\mathbb{N}_{\geq 1}$ are positive integers 
\end{definition}

\begin{definition}[Derived Resource Log]
    Let $\mi{EL}=(E,\mi{Att},\pi)$ be an event log and $\mathcal{CT}$, $\mathcal{AT}$ and $\mathcal{TT}$ be the collections of case types, activity types, and time types, respectively. 
    The resource log derived from $\mi{EL}$, $\mathcal{CT}$, $\mathcal{AT}$, and $\mathcal{TT}$ is
    % Previous definition (not using a bag notation)
    $\mi{RL}_{(\mi{EL},\mathcal{CT},\mathcal{AT},\mathcal{TT})}=(\mi{RE}_{(\mi{EL},\mathcal{CT},\mathcal{AT},\mathcal{TT})},m)$ where
		$\mi{RE}_{(\mi{EL},\mathcal{CT},\mathcal{AT},\mathcal{TT})}$ = 
		  $\{(r,ct,at,tt) \mid e\in E_{\mi{res}}
        \land \pi_{\mi{res}}(e) = r \in  \mathcal{R}
        \land \pi_{\mi{case}}(e) \in ct \in  \mathcal{CT}
        \land \pi_{\mi{act}}(e) \in at \in  \mathcal{AT}
        \land \pi_{\mi{time}}(e) \in tt \in  \mathcal{TT}\}$ is the set of derived resource events from $\mi{EL}$, $\mathcal{CT}$, $\mathcal{AT}$,  and $\mathcal{TT}$.
\end{definition}
\fi % ------------------- Definitions using a counter instead of a multiset notation ---------------

Table~\ref{tab:example_res_log} shows an example of a resource log derived from the sample event log in Table~\ref{tab:example_event_log} and some defined collections of case types, activity types, and time types, using the corresponding functions $\varphi_\mi{case}, \varphi_\mi{act},$ and $\varphi_\mi{time}$, respectively (see Def.~\ref{def:cat_types}). Below we include a more detailed explanation.
%\todo{More explanation}

\begin{itemize} 
	\item In Table~\ref{tab:example_event_log}, the \textit{customer type} event attribute has two values: \textit{normal} and \textit{VIP}. These are used to group cases into two categories and hence the two case types in Table~\ref{tab:example_res_log}. Normal cases include those identified as~654423 and 654424, while the VIP type contains case~654425. Using $\varphi_\mi{case}$, we can write  
	$\{ 654423, 654424 \} \subseteq \varphi_\mi{case}(\text{normal})$ and $\{ 654425 \} \subseteq \varphi_\mi{case}(\text{VIP})$.
	
	\item Four activity types are defined: \textit{register}, \textit{contact}, \textit{check}, and \textit{decide}. The seven values for activity names in Table~\ref{tab:example_event_log} are classified by these types as follows (denoted using $\varphi_\mi{act}$):
    $\{ \text{``register request''}, \text{``confirm request''} \} \subseteq \varphi_\mi{act}(\text{register})$,
    $\{ \text{``get missing info''}, \text{``pay claim''} \} \subseteq \varphi_\mi{act}(\text{contact})$, \\
    $\{ \text{``check insurance''} \} \subseteq \varphi_\mi{act}(\text{check})$, 
    $\{ \text{``accept claim''}, \text{``reject claim''} \} \subseteq \varphi_\mi{act}(\text{decide})$;
	
	\item The working hours of a day are divided into two timeframes, and hence lead to the two time types defined: \textit{morning} and \textit{afternoon}. Timestamps of events recorded in Table~\ref{tab:example_event_log} are classified accordingly, e.g., \\ $ \text{``30-08-2018 09:09''} \in \varphi_\mi{time}(\text{morning}) $, $ \text{``29-08-2018 15:02''} \in \varphi_\mi{time}(\text{afternoon}) $.
    \ifcomment
    time types = \{morning, afternoon\}, for which we have \\ $\{ \text{``30-08-2018 09:09''}, \text{``30-08-2018 09:22''}, \text{``30-08-2018 10:07''}, \\ \text{``30-08-2018 11:32''}, \text{``30-08-2018 11:45''} \} \subseteq \varphi_\mi{time}(\text{morning})$, \\ 
    $\{ \text{``29-08-2018 15:02''}, \text{``29-08-2018 16:08''}, \text{``29-08-2018 16:28''}, \\ \text{``30-08-2018 13:32''}, \text{``30-08-2018 14:09''} \} \subseteq \varphi_\mi{time}(\text{afternoon})$.
    \fi
\end{itemize}

% ----------------------------------------------------
\ifcomment
\begin{itemize} 
	\item case types = \{normal, VIP\}, for which we have \\ $\{ 654423, 654424 \} \subseteq \varphi_\mi{case}(\text{normal})$, $\{ 654425 \} \subseteq \varphi_\mi{case}(\text{VIP})$;
	\item activity types = \{register, contact, check, decide\}, for which we have \\ $\{ \text{``register request''} \} \subseteq \varphi_\mi{act}(\text{register})$, 
    $\{ \text{``get missing info''} \} \subseteq \varphi_\mi{act}(\text{contact})$, 
    $\{ \text{``check insurance''} \} \subseteq \varphi_\mi{act}(\text{check})$, 
    $\{ \text{``accept claim''}, \text{``reject claim''} \} \subseteq \varphi_\mi{act}(\text{decide})$;
	\item time types = \{morning, afternoon\}, for which we have \\ $\{ \text{``30-08-2018 09:09''}, \text{``30-08-2018 09:22''}, \text{``30-08-2018 10:07''}, \\ \text{``30-08-2018 11:32''}, \text{``30-08-2018 11:45''} \} \subseteq \varphi_\mi{time}(\text{morning})$, \\ 
    $\{ \text{``29-08-2018 15:02''}, \text{``29-08-2018 16:08''}, \text{``29-08-2018 16:28''}, \\ \text{``30-08-2018 13:32''}, \text{``30-08-2018 14:09''} \} \subseteq \varphi_\mi{time}(\text{afternoon})$.
\end{itemize}
\fi % ----------------------------------------------------

Each row in Table~\ref{tab:example_res_log} corresponds to a resource event, i.e., an event executed by some resource and characterized by an execution mode (a combination of some valid values from the above collections). There may be multiple occurrences of the same resource event. 
For example, event~(Pete, normal, register, afternoon) happened twice in Table~\ref{tab:example_res_log}, which means that Pete executed two events having the execution mode (normal, register, afternoon). 

\begin{table}[h!]
\caption{A fragment of some resource log (derived from the event log in Table~\ref{tab:example_event_log}).} 
%and~pre-defined case types, activity types, and time types}
\label{tab:example_res_log}
\centering
\begin{scriptsize}
    \begin{tabular}{cccc}
        \hline
        ~ \bf resource ~ &~ \bf case type ~ &  ~ \bf activity type ~ & ~ \bf time type ~ \\
        \hline
				... & ... & ... & ... \\
        Pete & normal & register & afternoon \\
        Pete & normal & register & afternoon \\
        Ann & normal & contact & afternoon \\
        John & normal & check  & morning \\
        Sue & normal & check  & morning \\
        Bob & VIP & register  & morning \\
        John & normal & decide & morning \\
        Sue & normal & decide & morning \\
        Mary & VIP & check & afternoon \\
        Mary & VIP & decide & afternoon \\
				... & ... & ... & ... \\
        \hline
    \end{tabular}
\end{scriptsize}
\end{table}

\subsection{Organizational Model}
\label{sec:framework/org_model}

We define the concept of an \emph{organizational model} (see Def.~\ref{def:org_model}) by incorporating the notion of execution modes. %It is a richer notion in that 
As such, the resulting model does not only define the grouping of resources ($\mi{RG}$ and $\mi{mem}$) as existing organizational models do, but also specifies the relationship between the resource groups and execution modes ($\mi{cap}$). It is important to introduce the latter as a novel component of an organizational model, in particular for a model discovered from event logs, since this component captures the knowledge of process execution relevant to the grouping of resources in the organization, and thus establish the linkage between organizational models and processes. 
%links resources via resource groups to execution modes in the close context of process execution. 

\begin{definition}[Organizational Model] \label{def:org_model}
    Let $\mathcal{R}$ be a set of resources and $\mathcal{CT}$, $\mathcal{AT}$ and $\mathcal{TT}$ be collections of case types, activity types, and time types.
    $\mi{OM}=(\mi{RG},\mi{mem},\mi{cap})$ is an organizational model where
        $\mi{RG}$ is a set of resource groups,
        $\mi{mem} \in \mi{RG} \rightarrow \powerset(\mathcal{R})$ maps each resource group onto its members,
        $\mi{cap} \in \mi{RG} \rightarrow \powerset(\mathcal{CT} \times \mathcal{AT} \times \mathcal{TT})$ maps each resource group onto its possible execution modes. \qed
\end{definition}

\begin{figure}[b!]
\centering
    \includegraphics[width=3.5in]{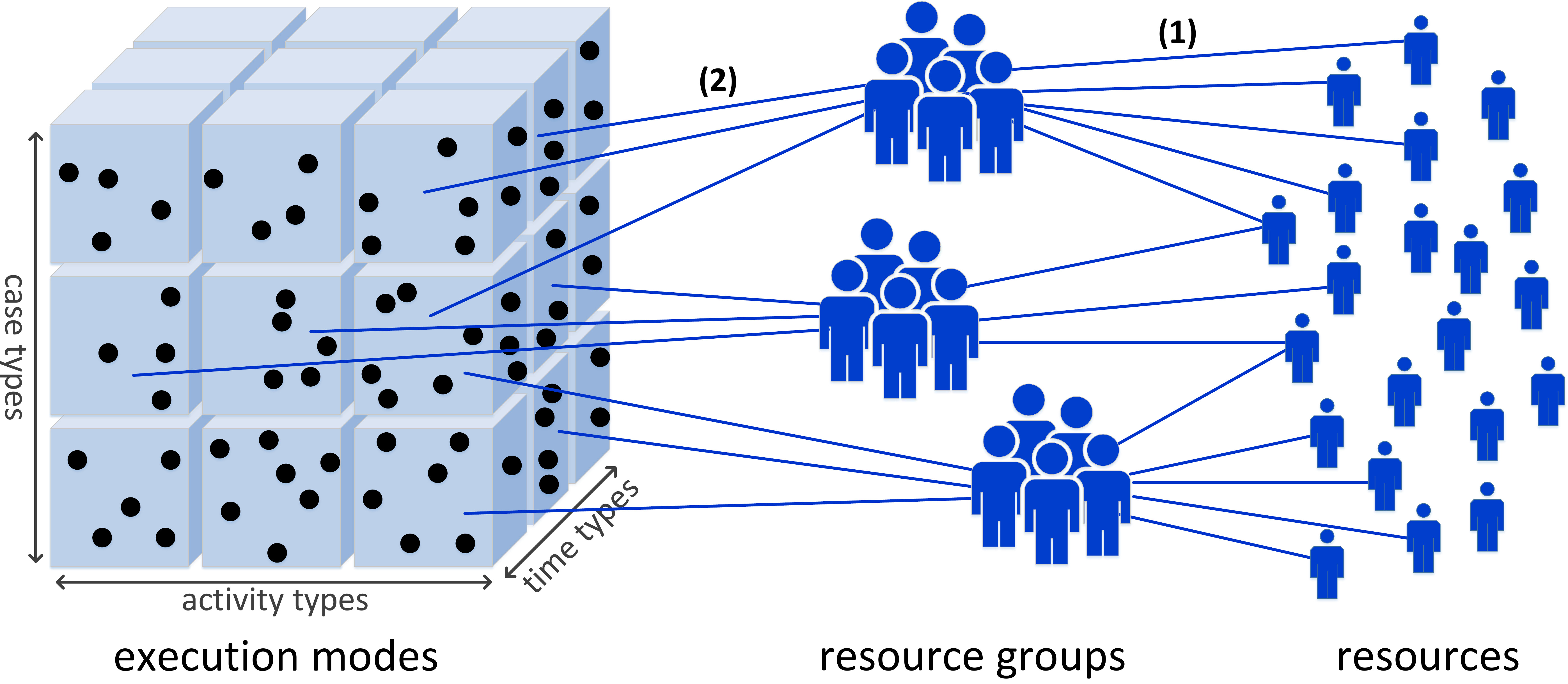}
\caption{Graphical view of the organizational model specified in Def.~\ref{def:org_model} which has a many-to-many relationship: (1) between resource groups and resources, and (2) between resource groups and execution modes.}
\label{fig:org-model}
\end{figure}

Fig.~\ref{fig:org-model} provides a graphical illustration of the above organizational model. 
The many-to-many relationships are used to capture the fact that resource groups may be overlapping in terms of members or execution modes, i.e., there may be two distinct groups $\mi{rg}_1$ and $\mi{rg}_2$ such that $\mi{mem}(\mi{rg}_1)\cap\mi{mem}(\mi{rg}_2)\neq\emptyset$ or $\mi{cap}(\mi{rg}_1)\cap\mi{cap}(\mi{rg}_2)\neq\emptyset$. 
%The same resource may be in multiple groups and the same execution mode may be supported by multiple groups.

\begin{figure}[t!]
\centering
    \includegraphics[width=\linewidth]{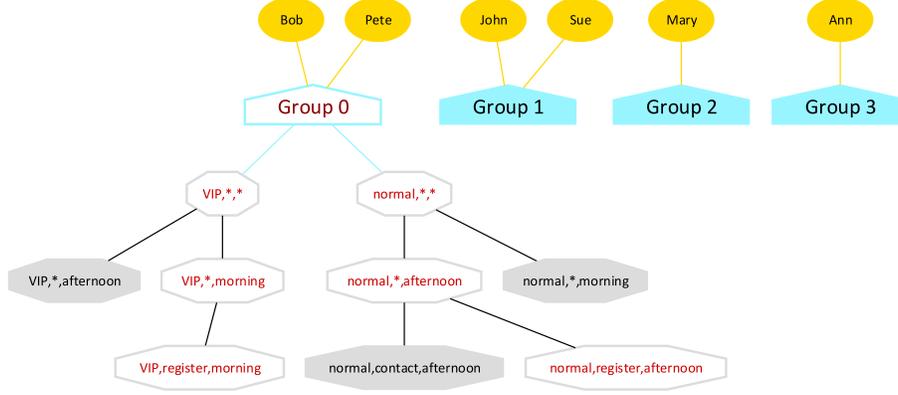}
\caption{Visualization of an example organizational model related to the sample event data in Table~\ref{tab:example_event_log} and its derived resource log in Table~\ref{tab:example_res_log}. 
Note that for a presentation purpose, only the capabilities of ``Group $0$'' are shown in the figure as octagons with red highlighted texts.}
%for better illustration here.}
\label{fig:example_org_model}
\end{figure}

Fig.~\ref{fig:example_org_model} shows an example organizational model with six resources as members of four resource groups, respectively. The capabilities of the first group (``Group $0$'' with member resources Bob and Pete) are illustrated as a tree of octagons with red highlighted texts, which shows that this group is capable of executing modes (VIP, register, morning) and (normal, register, afternoon).

\subsection{Conformance checking of Organizational Models} \label{sec:framework/conformance}
As discussed in our review of the related work, it remains an open issue that how to evaluate a discovered organizational model against the input event log. In this section, we address this issue by introducing \textit{conformance checking} based on the above richer notion of organizational models.
%Introducing the above richer notion of organizational models, which capture both resource grouping and the relevant knowledge of process execution, makes it possible to evaluate these models via \textit{conformance checking}~\cite{vanderaalst2016process}. 
Conformance checking refers to comparing modeled behavior with actual behavior (of process execution) when both a model and an event log are given, and can be applied for measuring the effectiveness of model discovery methods.
%While conformance checking is a well-acknowledged idea established in the broader process mining field, it is rarely concerned in the research on organizational model mining. To our knowledge, only the work by Baumgrass et al.~\cite{baumgrass2012conformance} directly addresses the idea, in which the authors prpose an approach that check the potential breaches of some given Role-Based Access Control (RBAC) policies using event log data. Yet their work does not provide the means for \ldots \todo{}
Below, we present the \textit{global conformance measures} for evaluating the degree of conformance between an organizational model and the corresponding event log, and also the \textit{local diagnostic measures} for the purpose of further investigating the reasons behind discrepancies between a model and an event log. 
%We propose the evaluation of organizational models 

%Business process conformance checking (a.k.a. conformance checking for short) is a family of process mining techniques to compare a process model with an event log of the same process.[1] It is used to check if the actual execution of a business process, as recorded in the event log, conforms to the model and vice versa.
%Bringing the execution modes component into the proposed definition of organizational models builds up the linkage between resource grouping and business processes, hence allows conformance checking between an organizational model and event log data, i.e., comparing modeled behavior with actual behavior in event logs.
%In the following, we introduce some notions along with measures that can be applied for checking the conformance of an organizational model with respect to an event log.

\subsubsection{Global conformance measures}
% How much are the commonalities/discrepancies between the behavior of a model and a log?
%Next, we propose two quality measures, \emph{fitness} and \emph{precision}, for evaluating organizational models discovered from event logs. They are used to compare a (discovered) organizational model with an (input) event log, which is often known as conformance checking in the field of process mining.

%Below, we discuss the two measures in detail. 
%has a good precision comparing with the log, and hence 
%($\mi{OM}$)  ($\mi{EL}$) %These are underpinned by quality measures in data mining, and \emph{tailored or adapted?} for conformance checking. 

These include \textit{fitness} and \textit{precision}, which can be used to measure the degree of commonalities or discrepancies between a model and an event log from two different perspectives.

% --- Fitness: event level (r, c, a, t) --- %

%\noindent\emph{Fitness.} 
%\paragraph{Fitness} Given an event log and an organizational model (such as one being discovered from the log), we introduce the notion of \emph{conforming events} to refer to those events that are both observed in the log and allowed in the model. If most of the events in the log are conforming events, it indicates a good fitness between the model and the log.

\paragraph{Fitness} 
Given an event log and an organizational model, \textit{fitness} considers the proportion of behavior in the log possible according to the model. 
%Here we measure fitness at the event level and 
To quantify fitness, we first introduce the notion of \emph{conforming events} (Def.~\ref{def:conf_events}). 
An event in a log is regarded as conformed with a given model, if its originating resource is allowed by the model to execute the event.
In light of this, we define a fitness measure at the event level (Def.~\ref{def:fitness}), which yields a value between $0$ and $1$. Note that only events with resource information ($E_{\mi{res}}$) should be considered.

\begin{definition}[Conforming Events] \label{def:conf_events}
    Let $\mi{EL}=(E,\mi{Att},\pi)$ be an event log and $\mi{OM}=(\mi{RG},\mi{mem},\mi{cap})$ an organizational model.
    $E_{\mi{conf}} = \{ e \in E_{\mi{res}} \mid \exists_{\mi{rg}\in \mi{RG}}\
    \pi_{\mi{res}}(e) \in \mi{mem}(\mi{rg}) \land
    \exists_{(\mi{ct},\mi{at},\mi{tt})\in \mi{cap}(\mi{rg})}\
    e \in [E]_{(\mi{ct},\mi{at},\mi{tt})}
    \}$ are all conforming events.    
    $E_{\mi{nconf}} =  E_{\mi{res}} \setminus E_{\mi{conf}}$ are all non-conforming events.
    \hfill\qed
\end{definition}

\begin{definition}[Fitness] \label{def:fitness}
    The fitness of an organizational model $\mi{OM}$ w.r.t. an event log $\mi{EL}$ is 
	\mbox{$\mi{fitness}(\mi{EL},\mi{OM}) = \frac{\card{E_{\mi{conf}}}}{\card{E_{\mi{res}}}}$}.
	\hfill\qed
\end{definition}

%The fitness measure yields a value between $0$ and $1$.
The fitness between a model and a log is good when most events in the log are conformed.
$\mi{fitness}(\mi{EL}, \mi{OM}) = 1$ if resources only performed events in~$\mi{EL}$ that they were allowed to perform according to~$\mi{OM}$.
$\mi{fitness}(\mi{EL}, \mi{OM}) = 0$ if no event in~$\mi{EL}$ was executed by a resource actually allowed to perform it according to~$\mi{OM}$. 
Following the definition, all events executed by a resource (see Def.~\ref{def:event_attributes}) in the example event log are conforming events. 
Hence, the example organizational model shown in Fig.~\ref{fig:example_org_model} has a fitness of $1$.
%\footnote{Events without resource information are ignored, and this is not a limitation as alternatively one can always add a resource ``nobody'' to the corresponding events in an event log.}. 

%\noindent\emph{Precision.}
%\paragraph{Precision} Given an event log and an organizational model, we introduce the notion of \emph{candidate resources} to refer to those resources that are allowed in the model to perform an event or a set of events in the log. 
%We propose that if there are usually very few candidate resources associated with an event in the log, it indicates a good precision between the model and the log. 
%Since each event is performed by at most one resource according to the log, precision is poor when all events can be done by all resources according to the model. 
%each event can only be performed by very few resources according to the model, it indicates a good precision between the model and the log.

\paragraph{Precision} 
Given an event log and an organizational model, \textit{precision} considers the extent of behavior allowed by the model with respect to the log.
To quantify precision, we propose the notion of \emph{candidate resources} (Def.~\ref{def:event_cand}).
For an event in a given log, its candidate resources with respect to a model refer to those resources allowed by the model to perform the event. 
The idea is that a perfectly precise model would only allow limited behavior and would never go beyond those recorded in a referenced log. In other words, it is assumed that an event can only be performed by a few candidate resources.

\begin{definition}[Candidate Resources] \label{def:event_cand}
    Let $\mi{EL}=(E,\mi{Att},\pi)$ be an event log and $\mi{OM}=(\mi{RG},\allowbreak \mi{mem},\mi{cap})$ an organizational model.
    $\mi{cand}:E\rightarrow\powerset(\mathcal{R})$ maps an event onto a set of candidate resources (in which an empty set is~possible). For each $e\in E$,  
    $\mi{cand}(e) = \{ r \in \mathcal{R} \mid  
    {\exists_{\mi{rg}\in \mi{RG}} \ r \in \mi{mem}(\mi{rg})} \land
    \exists_{(\mi{ct},\mi{at},\mi{tt})\in \mi{cap}(\mi{rg})} \ 
    e \in [E]_{(\mi{ct},\mi{at},\mi{tt})} \}$ is the set of candidate resources for event~$e$; and $\mi{cand}(E) = \\ \bigcup_{e\in E} \mi{cand}(e)$ is the overall set of candidate resources.
    \hfill\qed
\end{definition}

We also introduce the notion of \emph{allowed events} (Def.~\ref{def:allowed_events}). 
For an event in a given log, it is regarded as allowed by a model if there exists at least one candidate resource in the model to perform the event. 
%Note that, if a candidate resource of an event is the originating resource of the event, then such an event is both a conforming event and an allowed event. 
%A set of allowed events can be used to specify the behavior described by a model with respect to the corresponding event log. 
%The reason of specifying allowed events is to avoid accounting for behavior which is \textit{not} described by a model. 
%which refer to events in a log that could happen according to a given model (Def.~\ref{def:allowed_events}), in order to avoid accounting for behavior which is \textit{not} described by the model. 
%it is an allowed event when there exists at least one candidate resource in a model able to perform it, and is conforming only when it is both allowed and the actual performer is a candidate. 

\begin{definition}[Allowed Events] \label{def:allowed_events}
    Let $\mi{EL}=(E,\mi{Att},\pi)$ be an event log and $\mi{OM}=(\mi{RG},\mi{mem},\mi{cap})$ an organizational model. %$E_\mi{allowed} = \{ e \in E_{\mi{res}} \mid \exists_{\mi{rg}\in \mi{RG} \land (\mi{ct},\mi{at},\mi{tt})\in \mi{cap}(\mi{rg})}{e \in [E]_{(\mi{ct},\mi{at},\mi{tt})}} \}$ 
    $E_\mi{allowed} = \{e\in E_{\mi{res}}\mid\mi{cand}(e)\neq\emptyset\}$
    are all allowed events. 
    \hfill\qed
    %in the model within the scope of the set of events (executed by a resource) in the log. 
\end{definition}

Based on the above, the precision with respect to a model and an event log can be measured by considering the fraction of resources allowed by the model to perform the events in the log (Def.~\ref{def:precision}).
Like the fitness measure, the defined precision measure also yields a value between $0$ and $1$. 

\begin{definition}[Precision] \label{def:precision}
    The precision of an organizational model $\mi{OM}$ w.r.t. an event log $\mi{EL}$ is
		\mbox{$\mi{precision}(\mi{EL},\mi{OM})=\frac{1}{\card{E_\mi{allowed}}}\sum\limits_{e\in E_\mi{conf}}\frac{\card{\mi{cand}(E)}-\card{\mi{cand}(e)}+1}{\card{\mi{cand}(E)}}$}. 
		\qed
\end{definition}

$\mi{precision}(\mi{EL}, \mi{OM}) = 1$ if each event in $\mi{EL}$ allowed by $\mi{OM}$ is a conforming event, and has no other candidate but exactly \textit{the resource} who executed the event. %Put another way, 
On the contrary, $\mi{precision}(\mi{EL}, \mi{OM}) = 0$
%if each conforming event can be executed by all resources, or 
if none of the allowed events is a conforming event, i.e., no resource is allowed to perform events which it actually originated.
For instance, according to the organizational model in Fig.~\ref{fig:example_org_model}, the first event in the example log shown in Table~\ref{tab:example_event_log}, which is presented as the row ``654423, register request, 29-08-2018 15:02, Pete, normal'', has two candidate resources, Bob and Pete. Also, all the events executed by a resource in the log are allowed events. The precision value of this model with respect to the log is $0.883$, suggesting the model allows some extra behavior to happen, which was not observed according to the event log data.

% ------------------------------------ old definitions: not specific to a model
\ifcomment
\begin{definition}[Candidate Resources]
    Let $\mi{EL}=(E,\mi{Att},\pi)$ be an event log and $\mi{OM}=(\mi{RG},\allowbreak \mi{mem},\mi{cap})$ an organizational model.
    $\mi{cand} \in E \rightarrow \powerset(\mathcal{R})$ maps an event onto a set of candidate resources (in which an empty set is possible). % old definitions: not specific to a model
    $\mi{cand}(e) = \{ r \in \mathcal{R} \mid  \exists_{\mi{rg}\in \mi{RG}}\ 
    r \in \mi{mem}(\mi{rg}) \land
    \exists_{(\mi{ct},\mi{at},\mi{tt})\in \mi{cap}(\mi{rg})}\
    e \in [E]_{(\mi{ct},\mi{at},\mi{tt})} \}$.
    $\mi{cand}(E) = \bigcup_{e\in E} \mi{cand}(e)$ is the overall set of candidate resources.
\end{definition}

NOTE: Proposal~2 has been selected.

[DISCUSSION] 
Below are two proposals of how to define \emph{precision} measure, and we will need to decide on one.

\noindent Proposal~1: 

\begin{definition}[Precision/rc-measure] \label{def:rc-measure}
    The precision of an organizational model $\mi{OM}$ w.r.t. an event log $\mi{EL}$~is \\ %defined as follows: \\
		$\mi{precision}(\mi{EL},\mi{OM}) = \frac{1}{\card{E_\mi{conf}}} \sum\limits_{e \in E_\mi{conf}}$ $\frac{\card{\mi{cand}(E)} - \card{\mi{cand}(e)}}{\card{\mi{cand}(E)}-1}$.
\end{definition}

Key points: 
\begin{itemize}
	\item [1.] For precision, only fitting events $E_\mi{conf}$ are considered and all other events are discarded. 
	\item [Q] \emph{Is it a good idea to introduce dependency on fitness in the definition of precision? 
							Note that definitions of recall and precision in data mining are orthogonal and do not consider such dependency.} \\  
	\item [2.] Hence, precision should only be considered when fitness is good. %(This holds for both proposals)						
	\item [C] \emph{For most cases, yes, but there are exceptions (see two scenarios below)} 
	\begin{itemize}
			\item [S1] $\mi{EL}$ has 100 events (with resource information), 50 events are allowed in $\mi{OM}$, hence $\mi{fitness} = 0.5$; $\mi{OM}$ allows only these 50 events (and no other events), where each event has exactly one candidate resource, hence $\mi{precision} = 1$. 
			\item [S2] $\mi{EL}$ has 100 events (with resource information), 50 events are allowed in $\mi{OM}$, hence $\mi{fitness} = 0.5$; $\mi{OM}$ allows \textbf{80} events including the 50 conforming events, where each event has exactly one candidate resource, hence $\mi{precision} = 1$. 	
	\end{itemize}
	\item [Q] \emph{S1 and S2 have the same precision, should they? 
	                 For S1, should precision not be considered because of low fitness?} \\  
	\item [3.] $\mi{precision}(\mi{EL},\mi{OM}) = 0$ if each fitting event could be executed by all resources, i.e., the organizational model does not constrain the process. 
	\item [4.] $\mi{precision}(\mi{EL},\mi{OM}) = 1$ if each fitting event could only be executed by the resource that executed it.
	\item [C] \emph{Yes, given that the precision is only measured within the scope of fitting events.} \\ 
	\item [5.] It is assumed that there are at least two resources ($\card{\mi{cand}(E)}$>1). Hence, the case of one resource needs to be defined separately. 
	\item [Q] \emph{Would it be better to relax such assumption? I.e. not to differentiate the case of one resource with the rest?}
\end{itemize}

Key points: 
\begin{itemize}
	\item [1.] It considers allowed behaviour (instead of conforming behaviour), which aligns better with the definition of precision in data mining. 
	\item [2.] The relation $E_\mi{conf}\subseteq E_\mi{allowed}\subseteq E_\mi{res}$ holds, where $E_\mi{allowed} = E_\mi{conf}$ if only conforming events (and no other events) are allowed in $\mi{OM}$. 
	\item [3.] An allowed event may be a conforming event or a non-conforming event. For a conforming event, the best value is 1 and the worst value is $\frac{1}{\card{\mi{cand}(E)}}$. For a non-conforming event, the value is 0 and hence it is not shown in the formula. 
	\item [4.] It assumes at least one resource (instead of two), which is a bit more general.  
	\item [5.] Test on the two scenarios S1 and S2.
	\begin{itemize}
			\item [S1] $\mi{EL}$ has 100 events (with resource information), 50 events are allowed in $\mi{OM}$, hence $\mi{fitness} = 0.5$; $\mi{OM}$ allows only these 50 events (and no other events), where each event has exactly one candidate resource, hence $\mi{precision} = 1$. 
			\item [S2] $\mi{EL}$ has 100 events (with resource information), 50 events are allowed in $\mi{OM}$, hence $\mi{fitness} = 0.5$; $\mi{OM}$ allows \textbf{80} events including the 50 conforming events, where each event has exactly one candidate resource, hence $\mi{precision} =0.625$. 	
	%to check the fitness measure first. This holds for both proposals.} 	
	\end{itemize}
	If consider a F-measure, using this precision will also give a more reasonable result.  
\end{itemize}
\fi
% ------------------------------------ END old definitions: not specific to a model

%\emph{Let's carefully examine this measure. My understanding is:}
%\begin{itemize}
%	\item For a conforming event $e$, $\mi{cand}(e)$ returns at least one resource
%	\item $\mi{cand}(E)$ should be $\mi{cand}(E_\mi{res})$ (otherwise $\mi{cand}(E)$ also return resource candidates for those events without resource information)  
%	\item $\mi{cand}(E_\mi{res})$ counts for both conforming and non-conforming events that have resource information, while $\mi{cand}(e)$ only applies to conforming events. When the fitness is low, the number of the resource candidates returned by non-conforming events will then reduce the value of the measure). 
%\end{itemize}

\subsubsection{Local diagnostic measures}

While the global conformance measures provide an overall evaluation of the conformance between a model and an event log, they alone will not suffice when there is a need to %an interest in 
% one wishes to not only measure the extent of conformance but also in 
investigate the reasons behind non-conformance. Another offering of conformance checking is the local diagnostic analysis, which can reveal discrepancies in models with reference to the reality recorded by event logs~\cite{vanderaalst2016process}. The results may then inform improvements or corrections to be taken. 
Here, we introduce four possible measures that can be used to support local diagnostics on an organizational model given an event log.
%... [NEED REPHRASE] that can assist in providing insightful understanding towards resources' working performance as in groups from a process mining viewpoint.

\begin{itemize}
    \item \textit{Group relative focus} (on a given type of work) specifies how much a resource group performed this type of work compared to the overall workload of the group. It can be used to measure how the workload of a resource group is distributed over different types of work, i.e., work diversification of the group.
    
    \item \textit{Group relative stake} (in a given type of work) specifies how much of this type of work was performed by a certain resource group among all groups. It can be used to measure how the workload devoted to a certain type of work is distributed over resource groups in an organizational model, i.e., work participation by different groups.
    
    \item \textit{Group coverage} with respect to a given type of work specifies the proportion of members of a resource group that performed this type of work. %It can be used to measure 
    % an execution mode measures the ratio of group members who performed a certain type of work, to the whole group, i.e., work exposure of resources in this group.
    
    \item \textit{Group member contribution} of a member of a resource group with respect to a given type of work specifies how much of this type of work by the group was performed by the member. It can be used to measure how the workload of the entire group devoted to a certain type of work is distributed over the group members.
\end{itemize}

%The following part introduces some possible measures that could assist in providing insightful understanding towards resources' working performance as in groups from a process mining viewpoint, and also to aid local diagnostic on an organizational model given an event log.

We refer to the above measures as local diagnostic measures and they can be formulated precisely in Def.~\ref{def:grf} to  Def.~\ref{def:gmc}. 
Note that different types of work can be represented by different execution modes. 
Thus, the workload devoted to a certain type of work can be specified by the number of events capturing the occurrences of the corresponding execution mode.
And the workload of a resource group can be specified by the total number of events originated by its group members.
%frequency of the corresponding execution mode. 

\begin{definition}[Group Relative Focus] \label{def:grf}
%    Let $\mi{EL}=(E,\mi{Att},\pi)$ be an event log and $\mi{OM}=(\mi{RG},\mi{mem},\mi{cap})$ an organizational model.
    Given event log $\mi{EL}=(E,\mi{Att},\pi)$ and %be an event log and  
    organizational model $\mi{OM}=(\mi{RG},\mi{mem},\mi{cap})$, 
    for any resource group $\mi{rg}\in\mi{RG}$, its relative focus on 
    execution mode $(\mi{ct},\mi{at},\mi{tt})$ can be measured by \\
    $\mi{RelFocus}(\mi{rg}, (\mi{ct},\mi{at},\mi{tt})) = 
        \frac{\card{\{ e \in [E_{\mi{res}}]_{(\mi{ct},\mi{at},\mi{tt})}
                ~\mid~\pi_{\mi{res}}(e) \in \mi{mem}(rg) \}}}
             {\card{\{ e \in \mi{E_{res}}~\mid~\pi_{\mi{res}}(e) \in \mi{mem}(rg) \}}}
    $.
    \hfill\qed
\end{definition}

\begin{definition}[Group Relative Stake] \label{def:grs}
%    Let $\mi{EL}=(E,\mi{Att},\pi)$ be an event log and $\mi{OM}=(\mi{RG},\mi{mem},\mi{cap})$ an organizational model.
    Given event log $\mi{EL}=(E,\mi{Att},\pi)$ and %be an event log and  
    organizational model $\mi{OM}=(\mi{RG},\mi{mem},\mi{cap})$, 
    for any resource group $\mi{rg} \in \mi{RG}$, its relative stake in execution mode $(\mi{ct},\mi{at},\mi{tt})$ can be measured by \\
    $\mi{RelStake}(\mi{rg}, (\mi{ct},\mi{at},\mi{tt})) = 
        \frac{{\card{\{ e \in [E_{\mi{res}}]_{(\mi{ct},\mi{at},\mi{tt})}
                ~\mid~\pi_{\mi{res}}(e) \in \mi{mem}(rg) \}}}}
             {\card{[E_{\mi{res}}]_{(\mi{ct},\mi{at},\mi{tt})} }}
    $.
    \hfill\qed
\end{definition}

\begin{definition}[Group Coverage] \label{def:gc}
    Given event log $\mi{EL}=(E,\mi{Att},\pi)$ and %be an event log and  
    organizational model $\mi{OM}=(\mi{RG},\mi{mem},\mi{cap})$, 
    for any resource group $\mi{rg}\in\mi{RG}$, 
    the proportion of the group members covered by execution mode $(\mi{ct},\mi{at},\mi{tt})$ can be measured by 
    %the coverage of an execution mode $(\mi{ct},\mi{at},\mi{tt})$ in terms of the group members can be measured by \\
    $\mi{Cov}(\mi{rg}, (\mi{ct},\mi{at},\mi{tt})) = 
        \frac{\card{\{ r \in \mi{mem}(\mi{rg})~\mid~ 
                \exists_{e \in [E_\mi{res}]_{(\mi{ct},\mi{at},\mi{tt})}} \pi_{\mi{res}}(e) = r \}}}
             {\card{\mi{mem}(\mi{rg})}}
    $.
    \hfill\qed
\end{definition}

\begin{definition}[Group Member Contribution] \label{def:gmc}
%    Let $\mi{EL}=(E,\mi{Att},\pi)$ be an event log and $\mi{OM}=(\mi{RG},\mi{mem},\mi{cap})$ an organizational model.
    Given event log $\mi{EL}=(E,\mi{Att},\pi)$ and %be an event log and  
    organizational model $\mi{OM}=(\mi{RG},\mi{mem},\mi{cap})$, 
    for any resource %of a resource group %in the organizational model 
    $r \in \mi{mem}(\mi{rg})$ where $\mi{rg} \in \mi{RG}$, its contribution as a member of resource group $\mi{rg}$ to execution mode $(\mi{ct},\mi{at},\mi{tt})$ can be measured by \\
    $\mi{MemContr}(r, \mi{rg}, (\mi{ct},\mi{at},\mi{tt})) = 
        \frac{\card{\{ e \in [E_{\mi{res}}]_{(\mi{ct},\mi{at},\mi{tt})}~\mid~ \pi_{\mi{res}}(e) = r \}}}
             {\card{\{ e \in [E_{\mi{res}}]_{(\mi{ct},\mi{at},\mi{tt})}~\mid~ \pi_{\mi{res}}(e) \in \mi{mem}(rg) \}}}
    $.
    \hfill\qed
\end{definition}

\paragraph{Example} 
Let us revisit the example organizational model presented in Fig.~\ref{fig:example_org_model}.
For resource group ``Group $0$'' and one of its capabilities (VIP, register, morning), we can calculate these local diagnostic measures as follows.
\begin{itemize}
    \item $\mi{RelFocus}(\text{``Group 0''}, \text{(VIP, register, morning)}) = 0.333$,
    \item $\mi{RelStake}(\text{``Group 0''}, \text{(VIP, register, morning)}) = 1.0$,
    \item $\mi{Cov}(\text{``Group 0''}, \text{(VIP, register, morning)})      = 0.5$, and
    \item $\mi{MemContr}(\text{Bob}, \text{``Group 0''}, \text{(VIP, register, morning)}) = 0$,
    \item $\mi{MemContr}(\text{Pete}, \text{``Group 0''}, \text{(VIP, register, morning)}) = 1.0$.
\end{itemize}

The measured results imply that resources in ``Group 0'' consumed $33\%$ of their total workload ($\mi{RelFocus}$) carrying out work related to ``registering requests for VIP cases in the morning'', and this group is the only group that contributed to executing this type of work ($\mi{RelStake}$). Despite this, it is worth noting that only half of the group members ($\mi{Cov}$) have actually participated in conducting the work, i.e., resource Pete ($\mi{MemContr}$).

The example above shows how the use of local diagnostics may enable us to ``replay'' event data onto an organizational model and thus to study the actual behavior of resource groups and their members in process execution.
Utilizing these diagnostic results, we may explain the non-conformance between the organizational model and the event log implied by the imperfect model precision ($0.883$). After extending the examination to all resource groups and their capabilities using the diagnostic measures, we can show that ``Group 0'' is the only one in the example model that has a comparatively low group coverage considering all of its capabilities (as all other groups have a $1.0$ coverage over each of their capabilities).
It indicates that the non-conformance is located at the ``Group 0'' part in the model, which allows both Bob and Pete to be capable of executing the two modes. In the meantime, the event log does not provide any direct evidence showing such sharing of work between the two resources.

Nevertheless, note that further interpretation of the conformance checking results depends on the purpose of the model being whether descriptive or normative~\cite{vanderaalst2016process}. If the model is intended to be describing the observed event log data, then the revealed non-conformance may suggest that the model needs improvement, e.g., ``revising'' the model by dividing ``Group 0'' into two groups and linking each of them with only one execution mode. On the other hand, if the model is supposed to be normative, then such differences may lead to further analysis on resource performance. For instance, investigating why resource Bob and Pete did not process the registering of requests in a general way even when they are expected to do so.

\subsection{Organizational Model Mining Framework} 
\label{sec:framework/ommfw}

\begin{figure}[tb]
\centering
    \includegraphics[width=0.95\linewidth]{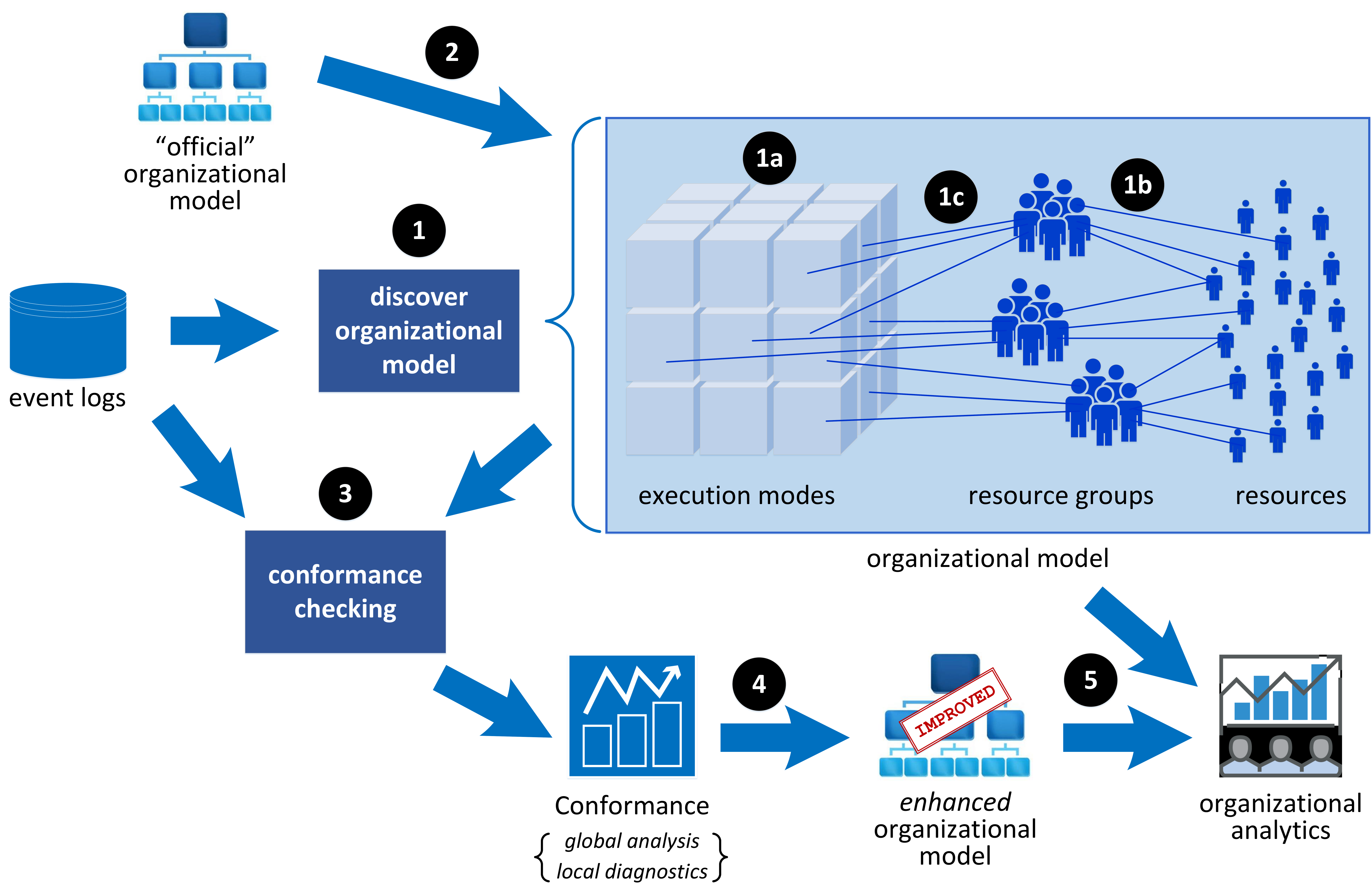}
\caption{Overview of the framework for organizational model mining.}
\label{fig:framework}
\end{figure}

Based on the notion of organizational models and the relevant conformance checking measures established above, we propose a framework for organizational model mining as depicted in Fig.~\ref{fig:framework}. The framework supports discovery, conformance checking, and extended analysis (e.g., analyzing the performance) of organizational models using event log data. 

As Fig.~\ref{fig:framework} illustrates, taking event log data as input, an organizational model can be discovered through step~(1). 
This consists of three sub-steps: 
(1a) learning execution modes, and
(1b) discovering resource groups, and
(1c) finding relationships between groups and execution modes.
Alternatively, an organizational model can be constructed manually, or an ``official'' organizational model is available and can be presented according to the structure proposed in Def.~\ref{def:org_model}, and these constitute step~(2). 
In step~(3), the conformance between the event log and organizational model can be checked using measures from a global perspective and a local perspective (as defined in Sect.~\ref{sec:framework/conformance}). 
The insights drawn from conformance checking can be used to revise or improve the ``as-is'' organizational model (i.e., step~(4)). 
Meanwhile, different organizational models can be used to simulate ``What if?'' scenarios (e.g., using relevant process data from the input event log), and it is also possible to repair a hand-made organizational model mediating between reality (as in the event logs) and the desired organization. These are referred to as organizational analytics (i.e., step~(5)). 

In the remaining of the paper, we focus on the discovery (step~(1)) and conformance checking (step~(3)) of organizational models. %and provide elaborate discussions in the next two subsections. 
It is worth noting that the framework proposed herein provides opportunities for addressing new research questions relevant to organizational model mining, which will be discussed as part of the future work at the end of the paper.

\section{Approach} 
\label{sec:approach}

\begin{figure}[ht!!!]
\centering
    \includegraphics[width=0.9\linewidth]{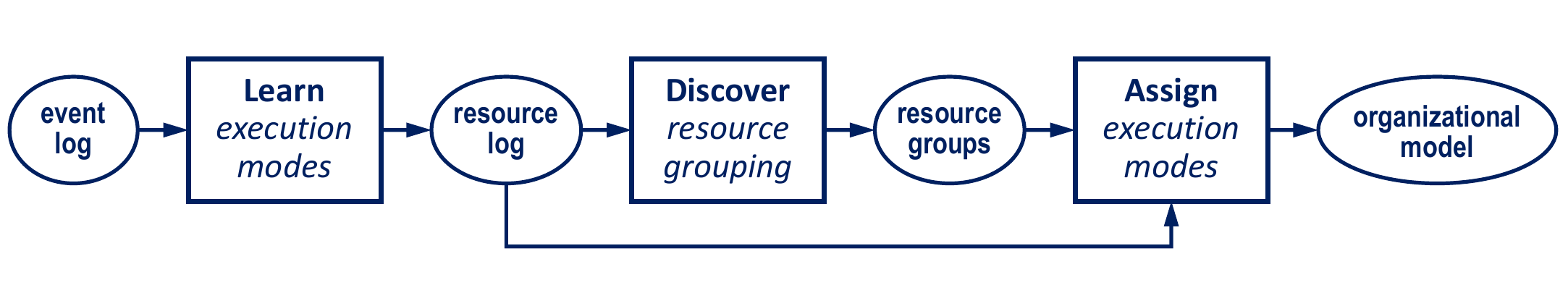}
\caption{An overview of the proposed approach.} 
\label{fig:approach}
\end{figure}

In this section, we introduce an approach underpinned by the framework proposed in the previous section, with a focus on discovering organizational models.
Fig.~\ref{fig:approach} depicts an overview of the approach. 
To begin with, an event log with the mandatory attributes and resource information (see Sect.~\ref{sec:framework/event_log}) is used as input for learning a set of execution modes (Phase~1). 
%using which one can derive as artifact a corresponding resource log. A resource log 
This yields as output a resource log, which is used to %then fed to the next phase of
discover resource grouping (Phase~2). %, e.g., project teams, departments, etc.
Discovered resource groups are then assigned a corresponding set of execution modes to describe the capabilities of their group members (Phase~3). An organizational model is produced  as the final artifact. The following subsections describe the methods and techniques used in these three phases, respectively.
%which incorporates several existing organizational model mining techniques. New methods are also presented to address challenges related to the discovery of a organizational model aligned with the proposed definition. 

\subsection{Learn Execution Modes} 
\label{sec:approach/phase1}

%A well-defined collection of such types would enable the partitioning of events, which helps organize events by different levels of granularity or dimensions and assists in the analysis of event data~\cite{vanderaalst2013process}. 
Learning execution modes requires identifying case, activity, and time types. This often depends on the available information recorded in the input event log as well as the purpose of analysis. 
For example, if an event log only records dates without the exact times when the events occurred, then defining time types at the scale of hours would be meaningless, whereas it may still be possible for a user to analyze daily or weekly process execution patterns using the same log.

A well-defined collection of execution modes %would enable the partitioning of events, which 
helps organize events at different levels of granularity or dimensions and assists in the analysis of event data~\cite{vanderaalst2013process}. %Building such a collection of types 
To propose a general method for learning execution modes is a challenging topic and will not be addressed in this paper. Instead, we present two simple methods as examples to show how execution modes can be derived from an event log. %rather na\"{i}ve 
%is out of the scope of this paper. 

\ifcomment
\begin{method*}[ATonly] \label{mtd:approach/phase1/at_only}
    Let $\mi{EL}=(E,\mi{Att},\pi)$ be an event log, define
    
    $\mathcal{CT} = \{\mathcal{C}\}$;
    $\mathcal{AT} = \{\{a\} \mid a \in \mi{range}(\pi_\mi{act})\}$;
    $\mathcal{TT} = \{\mathcal{T}\}$.   
\end{method*}
\fi

One method, namely \textsf{ATonly}, considers activity types only. 
%For example, the simplest solution is to let each of the activity labels recorded in an event log specifies a unique activity type. 
When such a method is applied, there are only one case type for all the cases and only one time type for all the timestamps recorded in the log, i.e., all events are regarded as having the same case type and time type. Since activity information is usually mandatory for event logs, \textsf{ATonly} can be applied in most situations and can serve as a starting point towards developing more advanced methods.
% i.e., the case and time dimensions are omitted when partitioning events into execution modes. 
%only the activity dimension is taken into account when partitioning events into execution modes. 

\ifcomment
\begin{method*}[CT+AT+TT] \label{mtd:approach/phase1/ct_at_tt}
    Let $\mi{EL}=(E,\mi{Att},\pi)$ be an event log, define
    
    $\mathcal{CT} = \{\mi{X} \in \mi{range}(\pi_\mi{case}) \mid
    \forall_{e_1, e_2 \in \mi{E}} \;
    \pi_{\mi{o}}(e_1) = \pi_{\mi{o}}(e_2) \land \pi_{\mi{case}}(e_1) \in \mi{X} \land \pi_{\mi{case}}(e_2) \in \mi{X}
    \lor
    \pi_{\mi{o}}(e_1) \neq \pi_{\mi{o}}(e_2)
    \}$, where $\mi{o}$ denotes a categorical case-level attribute;
    $\mathcal{AT} = \{\{a\} \mid a \in \mi{range}(\pi_\mi{act})\}$;
    $\mathcal{TT} = \{\{t\} \mid t \in \mi{range}(\mathscr{T})\}$, where $\mathscr{T}$ denotes a mapping from timestamps to values of a standard unit of time.
\end{method*}
\fi

The other method, namely \textsf{CT+AT+TT}, takes into account all three dimensions, i.e., case, activity, and time dimensions, when building execution modes. Case types may be derived according to different variants of cases specified by, e.g., case-level attributes recorded in event logs. 
Time types may be defined according to a predefined time unit. For example, selecting a time unit that refers to different days of a week (i.e., Monday, Tuesday, etc.) will result in seven time types for events recorded in an event log. 
%associated with Monday, Tuesday, etc.
%For any event, the corresponding time type will then be its timestamp mapped onto the selected time unit. 
%Meanwhile, defining time types can be done by using 
%based on the information that distinguishes between different variants of cases, 
% --- Move the use of trace clustering to the experiment section as an "idea of implementation of such a method":
%It is also possible to utilize results from applying trace clustering~\cite{bose2009} on an input event log to inform the inherent categorization of cases. 
%Meanwhile, activity names are still used for deriving activity types as in method \textsf{ATonly}.

Table~\ref{tab:example_exec_mode} shows the results of applying an example of \textsf{ATonly} method and an example of \textsf{CT+AT+TT} method, respectively, to the sample event log in Table~\ref{tab:example_event_log} to obtain the collection of execution modes. 
For \textsf{ATonly}, the simplest solution is used in which each distinct activity label recorded in the log is considered as a unique activity type. 
For \textsf{CT+AT+TT}, the case-level attribute ``customer type'' is selected, and the time unit is set to weekdays. Each event in the original log is now mapped onto an execution mode in Table~\ref{tab:example_exec_mode}.  
    
Once a collection of execution modes has been determined, a resource log can be derived. 
Compared to the original event log, a resource log provides a more focal view on event data from the resource perspective, and uses execution modes as resource attribute data. A derived resource log will be used as input to discover resource grouping in the next phase. 
% directly according to the definition. 
% of discovering an organizational model.

\begin{table}[b!]
\caption{Execution modes obtained after applying an \textsf{ATonly} method and a \textsf{CT+AT+TT} method, respectively, to the example event log in Table~\ref{tab:example_event_log}.} \label{tab:example_exec_mode}
\centering
\begin{threeparttable}
\begin{scriptsize}
    \begin{tabular}{l|l}
        \hline
        ~ \bf Execution modes  ~ & ~ \bf Execution modes ~ \\
        ~ \bf (applying \textsf{ATonly}) ~ & ~ \bf (applying \textsf{CT+AT+TT}) ~ \\
        \hline
        				... & ... \\
        ($\bot$, register request, $\bot$) & (normal, register request, Wednesday) \\ % Pete
        ($\bot$, register request, $\bot$) & (normal, register request, Wednesday) \\ % Pete
        ($\bot$, confirm request, $\bot$) & (normal, confirm request, Wednesday) \\ % (none)
        ($\bot$, get missing info, $\bot$) & (normal, get missing info, Wednesday) \\ % Ann
        ($\bot$, confirm request, $\bot$) & (normal, confirm request, Wednesday) \\ % (none)
        ($\bot$, check insurance, $\bot$) & (normal, check insurance, Thursday) \\ % John
        ($\bot$, check insurance, $\bot$) & (normal, check insurance, Thursday) \\ % Sue
        ($\bot$, register request, $\bot$) & (VIP, register request, Thursday) \\ % Bob
        ($\bot$, accept claim, $\bot$) & (normal. accept claim, Thursday) \\ % John
        ($\bot$, reject claim, $\bot$) & (normal, reject claim, Thursday) \\ % Sue
        ($\bot$, pay claim, $\bot$) & (normal, pay claim, Thursday) \\ % (none)
        ($\bot$, confirm request, $\bot$) & (VIP, confirm request, Thursday) \\ % (none)
        ($\bot$, check insurance, $\bot$) & (VIP, check insurance, Thursday) \\
        ($\bot$, accept claim, $\bot$) & (VIP, accept claim, Thursday) \\
        ($\bot$, pay claim, $\bot$) & (VIP, pay claim, Thursday) \\ % (none)
        				... & ... \\
        \hline
    \end{tabular}
    \begin{tablenotes}
        \item Note: $\bot$ denotes an undefined value on the corresponding dimension in the execution modes.
    \end{tablenotes}
\end{scriptsize}
\end{threeparttable}
\end{table}

\subsection{Discover Resource Grouping} \label{sec:approach/phase2}
%An organizational model describes both the structuring of individual resources into groups and the capabilities of these groups. We first discuss how to discover the grouping of resources.

%In this paper 
We refer to resource groups as the general concept of resource classes~\cite{dumas2018fundamentals} well-accepted in business process management, i.e., members in a resource group are interchangeable in terms of carrying out similar types of work in process execution. 
Such resource groups may correspond to various forms of grouping entities in organizations. %depending on different contexts.
For example, employees from the accounting department in a company share responsibilities in executing tasks related to the payment and receipt of bills rather than those of manufacturing products. Another example can be found in the use of customer teams where resources as members of a team would likely be focusing on specific sorts of cases distinguished by the customers who initiate the cases.

In this sense, the discovery of resource grouping can be made through clustering resources with similarities in terms of their behavior captured by a resource log (derived from the previous phase). %, by using a derived resource log as a reference. 
To this end, we first consider building a set of features to characterize the behavior of each individual resource, and do so by applying an idea inspired by the notion of ``performer-by-activity matrix''~\cite{vanderaalst2005discovering,song2008towards,appice2018towards,yang2018finding}. The features of a resource are determined by both the variety and the event frequency of execution modes that the resource has executed. Using the resource log as input, this will produce a matrix of which each row corresponds to the feature vector of a resource, and the values correspond to the frequency of this resource originating different execution modes. Table~\ref{tab:example_res_feature_matrix} shows the feature matrix resulting from applying the idea on a derived resource log, which is related to the execution modes learned using method \textsf{CT+AT+TT}, as shown in the previous example (see Table~\ref{tab:example_exec_mode}).

\begin{table}[b]
\caption{An example 6 $\times$ 7 resource feature matrix obtained, with respect to the execution modes learned from applying method \textsf{CT+AT+TT} (see Table~\ref{tab:example_exec_mode}).} \label{tab:example_res_feature_matrix}
\centering
\begin{scriptsize}
    \begin{tabular}{c|ccccccc}
        \hline
        ~ \bf Resource ~ & \multicolumn{7}{c}{\bf Features} \\
        \hline
        Ann & 0 & 1 & 0 & 0 & 0 & 0 & 0 \\
        Bob & 0 & 0 & 0 & 1 & 0 & 0 & 0 \\
        John & 0 & 0 & 1 & 0 & 1 & 0 & 0 \\
        Mary & 0 & 0 & 0 & 0 & 0 & 1 & 1 \\
        Pete & 2 & 0 & 0 & 0 & 0 & 0 & 0 \\
        Sue & 0 & 0 & 1 & 0 & 1 & 0 & 0 \\
        \hline
    \end{tabular}
\end{scriptsize}
\end{table}

With a resource feature matrix obtained, the clustering of resources can be done in a straightforward way employing well-established data mining techniques. Some previous studies on organizational model mining address the issue by adopting cluster analysis~\cite{song2008towards,yang2018finding} or community detection~\cite{appice2018towards,ferreira2011discovering} techniques.
Cluster analysis aims at grouping a set of data objects into multiple clusters such that objects within a cluster have high similarity but are dissimilar to those in other clusters~\cite{han2011data}. The idea of applying these cluster analysis techniques is straightforward with the resource feature matrix as input. 
%AHC (Agglomerative Hierarchical Clustering)~\cite{song2008towards} applies the hierarchical clustering algorithm; GMM (Gaussian Mixture Model)~\cite{yang2018finding} uses the classic Gaussian mixture model with EM (Expectation-Maximization) algorithm; MOC (Model-based Overlapping Clustering)~\cite{yang2018finding} adopts another probabilistic model to conduct the cluster analysis.
Community detection is the discovery of community structures in networks, which refers to the division of nodes into groups within which the connections are denser compared to those between groups~\cite{newman2004finding}. For methods employing community detection techniques, a resource ``social network'' first needs to be constructed, and then a community detection algorithm can be applied to discover communities in the network forming the resource groups. While building the network, each resource is considered a node, and the similarity among resource feature vectors in the matrix is measured to determine the pairwise connection. 
%MJA (Metrics based on Joint Activities)~\cite{song2008towards} uses a designated threshold to keep only the stronger connections in the network, then recognizes the connected components in the resulting network as resource communities; the online organizational unit discovery approach in \cite{appice2018towards} applies the state-of-art link partitioning algorithm on the network to discover resource communities.
Although a variety of choices of techniques that can fit in the current phase as solutions for discovering resource grouping, their effectiveness may vary from case to case, hence it is necessary to make a selection based on the characteristics of the input data, i.e., the resource feature matrix.

Finally, we discuss a couple of technical issues concerning the discovery of resource grouping.
The first one is to determine the number of groups to be identified, which depends on both the latent structuring of the resources and the aim of analysis. For most techniques, this is often done by specifying one or more parameters before running the algorithm, e.g., the number of distributions used, resolution, etc. In cluster analysis, methods like cross-validation and elbow method may help determine a suitable configuration of parameters and hence the number of groups to be identified.
Another issue concerns whether resource groups are disjoint from or overlapping each other. In modern organizations, it is common to have individuals holding multiple roles or positions and thus sharing membership across organizational groups. To discover a resource grouping in which a resource can belong to more than one group, overlapping clustering or overlapping community detection techniques are needed. Some existing work on the topic (e.g., \cite{appice2018towards, yang2018finding}) employs such techniques and supports identifying resource groups with overlaps.

\ifcomment
Table~\ref{tab:prev_work_discover} shows an overview of the methods used in previous work. 
\todo{Describe how the existing techniques could fit in the current phase.}

\begin{table*}[htbp]
\caption{An overview of the methods used in previous work to discover resource grouping.} \label{tab:prev_work_discover}
\centering
    \begin{tabular}{c|c|c|c}
        \hline
        ~ Category ~  & ~ Method ~ &  ~ Summary ~ & ~ Reference ~ \\
        \hline
        \multirow{3}{*}{Cluster analysis} & AHC (Agglomerative Hierarchical Clustering) & 
        
        & \cite{song2008towards} \\
        \cline{2-4}
        & GMM (Gaussian Mixture Model) & \ldots & \multirow{2}{*}{\cite{yang2018finding}} \\
        \cline{2-3}
        & MOC (Model-based Overlapping Clustering) & \ldots & \\
        \hline
        \multirow{2}{*}{Community detection} & MJA (Metrics based on Joint Activities) & \ldots & \cite{song2008towards} \\
        \cline{2-4}
        & Online organizational unit discovery & \ldots & \cite{appice2018towards} \\
        \hline
    \end{tabular}
\end{table*}
\fi

\subsection{Assign Execution Modes} \label{sec:approach/phase3}
To produce a final organizational model, the capabilities of discovered resource groups need to be determined. In the proposed approach, this is achieved by assigning each resource group a set of relevant execution modes with reference to the log. We first consider a simple strategy that assumes each historic execution carried out by a member of a resource group contributes to the group's capabilities. 
We refer to this strategy as method \textsf{FullRecall}. The idea behind this method is similar to that of Entity Assignment method in~\cite{song2008towards}. 

\begin{method*}[FullRecall]
    Let $\mi{rg} \in \mi{RG}$ be a resource group discovered from event log $\mi{EL}$, and $\mi{CT} \times \mi{AT} \times \mi{TT}$ be a collection of execution modes derived from $\mi{EL}$. 
    The capabilities of resource group~$\mi{rg}$ is  
    $\mi{cap}(\mi{rg}) = \\ \{
        (\mi{ct}, \mi{at}, \mi{tt}) \in \mi{CT} \times \mi{AT} \times \mi{TT} 
        \mid
        \exists_{\mi{r} \in \mi{mem}(\mi{rg})} (\mi{r}, \mi{ct}, \mi{at}, \mi{tt}) \in 
        \mi{RL}(\mi{EL}, \mi{CT}, \mi{AT}, \mi{TT})
    \}$.
\end{method*}

\textsf{FullRecall} assigns an execution mode to a discovered group as long as one of the group members has originated an event of this mode. As such, it ensures that all observed behavior recorded in a log is captured by the resulting organizational model.

%\todo{How to introduce the motivation of proposing yet another method for mode assignment?}
Nevertheless, applying such a strategy runs the risk of considering excessive modes as group capabilities. For example, execution modes carried out by only a few members would still be regarded as being shared among all resources within a group. 
Hence, we go a step further to consider distinguishing the most relevant execution modes from all the executions, and only assign these modes as group capabilities. This requires measuring the extent of relatedness of a mode with respect to a group, and we approach it under two assumptions:
\begin{itemize}
    \item the capabilities of a resource group should be reflected by the types of work in which the group had considerable participation (which can be measured by \textit{group relative stake}), and % with event logs given), and,
    \item the capabilities of a resource group should be reflected by the types of work which was carried out by most of the group members (which can be measured by \textit{group coverage}). %with event logs given).
\end{itemize}

%To this end, 
Based on the above, we present the following method, namely \textsf{OverallScore}, for assigning execution modes.  \textsf{OverallScore} considers two factors jointly, \textit{group relative stake} and \textit{group coverage}, by using a weighted average which enables a varying extent of focus to be set for different factors. 
%Instead, the other method \textsf{Overall-HM} implements the idea by using harmonic mean so that the relatedness score would not be dominated by only one but is balanced between the two aspects being measured.

\begin{method*}[OverallScore]
    Let $\mi{rg} \in \mi{RG}$ be a resource group discovered from event log $\mi{EL}$, and $\mi{CT} \times \mi{AT} \times \mi{TT}$ be a collection of execution modes derived from $\mi{EL}$. 
    Given a threshold $\lambda \in (0, 1]$, and weights $\omega_1, \omega_2 \in [0, 1], \omega_1 + \omega_2 = 1$,
    $\mi{cap}(\mi{rg}) = \{
        (\mi{ct}, \mi{at}, \mi{tt}) \in \mi{CT} \times \mi{AT} \times \mi{TT} 
        \mid
        \omega_1 \cdot \mi{RelStake}(\mi{rg}, (\mi{ct}, \mi{at}, \mi{tt}))
        + \\ \omega_2 \cdot \mi{Cov}(\mi{rg}, (\mi{ct}, \mi{at}, \mi{tt})) \geq \lambda
    \}$.
\end{method*}

\ifcomment
\begin{method*}[OverallScore-HM]
    Let $\mi{rg} \in \mi{RG}$ be a discovered resource group from event log $\mi{EL}$, $(\mi{ct}, \mi{at}, \mi{tt}) \in \mathcal{CT} \times \mathcal{AT} \times \mathcal{TT}$ be an execution mode. We have
    $(\mi{ct}, \mi{at}, \mi{tt}) \in \mi{cap}(\mi{rg})$, if
    $\frac{2 \cdot \mi{RelStake}(\mi{rg}, (\mi{ct}, \mi{at}, \mi{tt})) \cdot \mi{Cov}(\mi{rg}, (\mi{ct}, \mi{at}, \mi{tt}))}
    {\mi{RelStake}(\mi{rg}, (\mi{ct}, \mi{at}, \mi{tt})) + \mi{Cov}(\mi{rg}, (\mi{ct}, \mi{at}, \mi{tt}))} \geq \lambda$,
    where $\lambda \in (0, 1]$. 
\end{method*}
\fi

As a result, an execution mode will be assigned as part of a group's capabilities only when it is observed in the log that sufficient members in the group have a certain amount of participation in executing the mode. Note that other factors reflecting the relatedness between resource groups and execution modes can be considered and are not necessarily limited to the proposed ones. %Other ways for quantitative measures can also be developed respectively and be incorporated into the method.

Table~\ref{tab:example_mode_assignment} presents an example of applying the proposed methods \textsf{FullRecall} and \textsf{OverallScore} for execution mode assignment considering a group of two resources (namely John and Sue) and the execution modes shown in Table~\ref{tab:example_exec_mode} as input.

\begin{table}[tb]
    \caption{Execution mode assignment results obtained for a given resource group of two members applying methods \textsf{FullRecall} and \textsf{OverallScore} ($\omega_1 = \omega_2 = 0.5, \lambda = 0.8$), when method \textsf{CT+AT+TT} is used for learning execution modes (see Table~\ref{tab:example_exec_mode}).} \label{tab:example_mode_assignment}
    \centering
    \begin{scriptsize}
        \begin{tabular}{c|l|l}
            \hline
            ~ \bf \multirow{2}{*}{Members} ~ & ~ \bf Group capabilities ~ & ~ \bf Group capabilities ~ \\
            & ~ \bf (applying \textsf{FullRecall}) ~ & ~ \bf (applying \textsf{OverallScore}) ~ \\
            \hline
            ~ \multirow{3}{*}{``John'', ``Sue''} & (normal, check insurance, Thursday) & \multirow{3}{*}{(normal, check insurance, Thursday)} \\
            & (normal, accept claim, Thursday) & \\
            & (normal, reject claim, Thursday) & \\
            \hline
        \end{tabular}
    \end{scriptsize}
\end{table}

%% ------------------------ post-refinement
\ifcomment % 
It is worth noticing that given a resource group, applying either \textsf{AssignbyAll} or \textsf{AssignbyProportion} could result in an empty set being produced, i.e. no execution mode can be associated with this group. This may happen especially when the discovery of resource grouping (Phase 2.2) fails to generate a high quality clustering and incurs rather significant intra-group dissimilarities among member resources. We address such issue by introducing a post-clustering refinement procedure.

An invalid result is returned when there exists a group of of which all (or a certain proportion) members share no common execution history. In order to guarantee at least one execution mode can be linked with such a group, in the refinement procedure we split the group such that, with the least number of sub-groups to be produced, each of them is valid for assigning execution modes using \textsf{AssignbyAll}/\textsf{AssignbyProportion}.
\todo{Need to be extremely careful here since rigorous formulation and solution are required.}
We model the challenge as a classical set cover problem and provide a sub-optimal solution based on SetCoverGreedy algorithm. Note that the split is done separately for each invalid group; and any two groups that share exactly the same membership will be recognized as one identical group in the final results after the refinement procedure.

For the methods above, execution histories of all member resources contribute equally to the assignment of execution modes to the group, which poses a challenge for the phase of discovering resource groups.
The last method \textsf{AssignbyWeighting} considers only histories of some certain members. A set of the most ``representative'' member resources is first determined given a resource group, and the execution modes to be assigned are resulted from the union set of their execution histories. In this paper, the selection of representative resources is done by utilizing the output from the previous clustering phase, taking members with top-\textit{k} highest Local Outlier Factor (LOF)~\cite{breunig2000} scores as representatives of a group. 
As various other strategies can be adopted to specify the representative set, method \textsf{AssignbyWeighting} lends more possibilities to a user to incorporate the use of more external knowledge related to resources in determining the capabilities of resource groups.
\fi
%% ------------------------

\section{Implementation} \label{sec:implementation}

%\subsection{OrgMiner}
We developed a prototype which implements the presented approach and the conformance checking measures. The prototype consists of a Python library named OrgMiner\footnote{
    OrgMiner library. 
    \url{https://orgminer.readthedocs.io/}
},
and a tool\footnote{
    Tool for organizational model mining and visualization. 
    \url{https://orgminer.readthedocs.io/en/latest/examples/infsyst2020yang-arya.html}
}
which allows a user to perform organizational model mining with a given event log, and to visualize the details of a discovered model.

{OrgMiner} includes modules that provide the functionalities with respect to different phases in the approach. Users are allowed to select from several alternative methods for each phase, presented as follows.

\paragraph{Execution mode learning}
Three different methods are implemented, including
\begin{itemize}
    \item \textsf{ATonly}, which considers only the activity types and neglects the other two dimensions. Activity types are constructed such that each activity label defines a unique type;
    \item \textsf{CT+AT+TT (case attribute)}, which considers all of the three dimensions. Activity types are built based on activity labels as in \textsf{ATonly}, and time types are built by categorizing events into 7 classes based on the weekday of the timestamps.
    Users may specify a case-level attribute in a given event log, for which each distinct value will be recognized as a case type to be used;
    \item \textsf{CT+AT+TT (trace clustering)}, which also considers all the three dimensions, except that the construction of case types is based on the results of utilizing trace clustering~\cite{bose2009}. The case type of an event is determined by the trace cluster to which it belongs. The Context Aware Trace Clustering~\cite{bose2009} technique was adopted for implementation.
\end{itemize}

\paragraph{Resource grouping discovery} 
Two methods proposed in the previous research on organizational model mining were incorporated into the implementation, i.e., \textsf{AHC}~\cite{song2008towards} and \textsf{MOC}~\cite{yang2018finding}. Both methods require specifying a proximity measure (e.g., Euclidean distance) and the number of clusters to be derived.

\paragraph{Execution mode assignment}
The two methods proposed in Sect.~\ref{sec:approach/phase3} are both implemented. For method \textsf{OverallScore}, users are required to specify the values for the threshold and the weightings.

These alternative methods can be configured independently, and the final output will be obtained by systematically deploying the selected methods. %assembling a chain of the selected methods. 
This design provides %enables not only the 
flexibility when applying the approach and, more importantly, it supports the extensibility for integrating the implementations of the new methods in the future. %that realize the framework. 
%Additionally, it also offers the possibility of integrating the principles of scientific workflow into the experiments, which will be elaborated in the experimental design. 

\ifcomment
%\subsection{Interactive Demo}
We built a prototype tool with an interface\footnote{
    Prototype tool for organizational model mining. 
    \url{https://orgminer.readthedocs.io/en/latest/examples/infsyst2020yang-arya.html}
}
based on OrgMiner, which allows a user to perform organizational model mining by providing an event log and specifying the method options, and then to visualize the details of a discovered model.

An interactive demo application\footnote{
    Interactive demo for organizational model mining. 
    \url{https://orgminer.readthedocs.io/en/latest/examples/infsyst2020yang-arya.html}
}
was built upon OrgMiner, which allows a user to perform organizational model mining with an event log given, and then to visualize the details of the discovered model. 
%Model visualization figures as will be shown in the experiment results section were produced from using this demo.
%Figures of organizational model visualization shown in this paper were produced from using the demo.
\fi

\section{Evaluation} \label{sec:evaluation}
We have conducted extensive experiments on real-life event log data to evaluate the presented approach for discovering organizational models using the proposed conformance checking measures. In this section, we introduce the event log dataset adopted for evaluation, and describe the setup of experiments. Results are then illustrated and analyzed, followed by a discussion of findings. 

\subsection{Experiment Dataset}
Real-life data sets are adopted for experiments, including one event log sourced from the CoSeLoG project~\cite{buijs2014flexible} and one from the Business Process Intelligence Challenge (BPIC). These event log data are publicly available online at {4TU Centre for Research Data}\footnote{{4TU Centre for Research Data}. \url{https://data.4tu.nl/}
}.
The first log ({WABO}~\cite{buijs2014wabo}) records data from the receipt phase of a building permit process in an anonymous Dutch municipality as part of the CoSeLoG project.
The second ({BPIC17}~\cite{vandongen2017bpic}) contains event data related to a loan application process in a Dutch financial institute.
These logs satisfy the basic requirements for event attributes as defined in Sect.~\ref{sec:framework/event_log}. They contain case, activity, time, and resource information, and also carry several case-level attributes such as the purpose of a loan application. 

To obtain the experiment dataset, a preprocessing step was conducted on the original logs to remove redundant event data --- only events recording the completion of %their generating 
activities\footnote{Refer to the standard transaction model for activities in the IEEE XES Standard~\cite{ieee2016xes}.} are kept. The filtering guarantees that one actual execution of an activity in the process would be counted once only.
Table~\ref{tab:dataset} reports some basic statistics of the preprocessed event logs ready for experiments.

\begin{table}[ht]
\caption{Descriptive statistics of the preprocessed event logs for experiments.} \label{tab:dataset}
\centering
\begin{scriptsize}
    \begin{tabular}{c|cccc}%|c}
        \hline
        ~ \bf {Log} ~ & ~ \bf \#cases ~ & ~ \bf \#events ~ & ~ \bf \#activities ~ & ~ \bf \#resources ~ \\
        \hline
        {WABO}      & 1,434     & 8,577     & 27 & 48 \\
        \hline
        %{BPIC12}    & 13,087    & 164,506   & 23 & 68 \\
        {BPIC17}    & 31,509    & 475,306   & 24 & 144 \\
        \hline
    \end{tabular}
\end{scriptsize}
\end{table}

\subsection{Design and Setup}\label{sec:evaluation/setup}
\begin{figure*}[!t]
\centering
    \includegraphics[width=\linewidth]{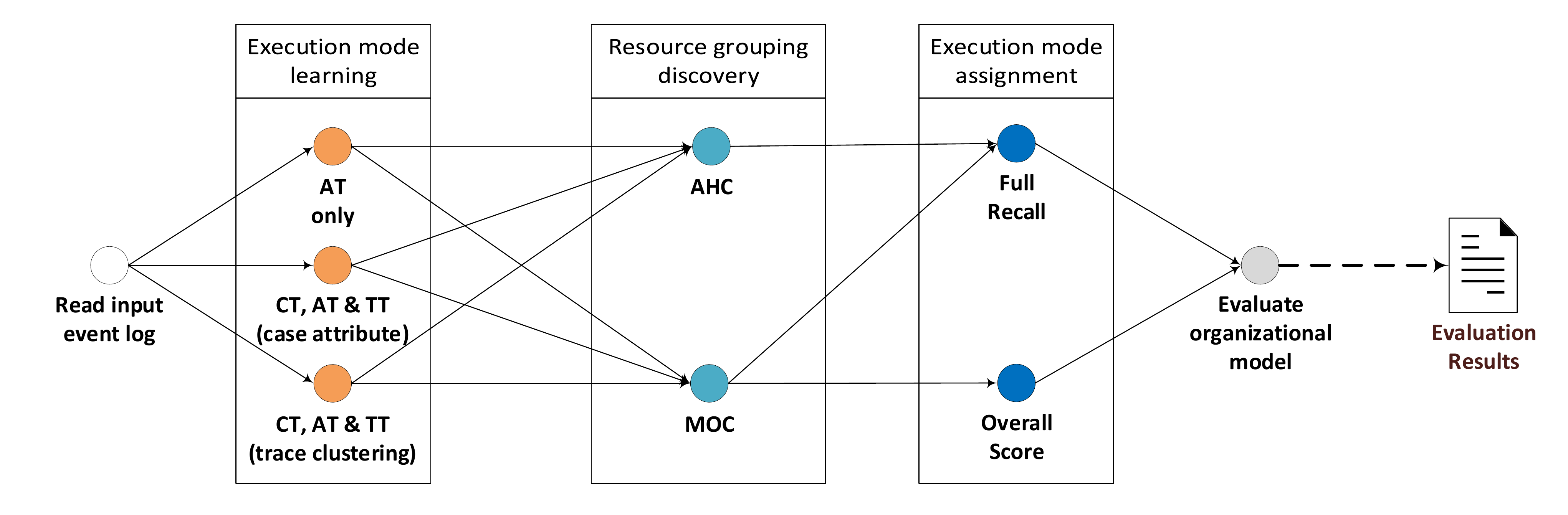}
\caption{An overview of the experiment settings. For an input event log, each path in the graph specifies a combination of methods when using the approach to discover an organizational model.}
\label{fig:sci_workflow}
\end{figure*}

The implemented approach includes several alternative methods at each intermediate phase (see Sect.~\ref{sec:implementation}).
In the experiments, we tested all the combinations of the alternatives in terms of model discovery and evaluation. Fig.~\ref{fig:sci_workflow} depicts an overview of our experiment settings, where a model is discovered with an execution mode learning method, a resource grouping discovery method, and a mode assignment method selected. 

The combination of all these alternatives resulted in a considerable number of organizational models to be discovered and evaluated for the two input logs. To address the issue and realize such an experiment design, we applied the principles of scientific workflow to automate the process of experiments by utilizing scripting and graph visualization software. This facilitated both the setup and conduct of our experiments while also benefiting future research in terms of sharing, replicating, and modifying the experiments.
The configuration of these methods\footnote{
    Experiment configuration. \url{https://orgminer.readthedocs.io/en/latest/examples/infsyst2020yang-replicate.html}
    } in the experiments are presented as follows.

\ifcomment
\begin{table}[tb]
\caption{Configuration of methods used in the experiments.} \label{tab:exp_config}
\centering
\begin{scriptsize}       
    \begin{tabular}{c|c|c}
        \hline
        ~ {\bf Phase} ~ & ~ {\bf Method} ~ & ~ {\bf Configuration} ~ \\
        \hline
        Execution & \textsf{ATonly} & (N/A) \\
        mode      & \textsf{CT+AT+TT (case attribute)} & \\
        \cline{2-3}
        learning  & \textsf{CT+AT+TT (trace clustering)} & Followed the configuration in the original paper~\cite{bose2009}. \\

    \end{tabular}
\end{scriptsize}
\end{table}
\fi

For execution mode learning methods, \textsf{ATonly} requires no extra configuration. 
For \textsf{CT+AT+TT (case attribute)}, we selected an attribute recording the channel of environmental permit application (with 5 possible categorical values) for log {WABO}, and the attribute recording the goal of the loan applied for by customers (with 14 possible categorical values) for log {BPIC17}.
For \textsf{CT+AT+TT (trace clustering)}, we followed the original configurations in the paper~\cite{bose2009} which proposes the trace clustering technique that we adopted for implementation (see Sect.~\ref{sec:implementation}).

Both methods for resource grouping discovery, i.e., \textsf{AHC} and \textsf{MOC}, applied the same configuration. The Euclidean distance was selected as the proximity measure. The issue of deciding the number of resource groups was addressed by employing the cross-validation technique~\cite{han2011data}, with the range of the potential number of groups set to $[2, 10]$.

As for execution mode assignment methods, \textsf{FullRecall} requires no extra configration. For the parameters used by method \textsf{OverallScore}, we performed grid search with the expected range of weighting values $\omega_1, \omega_2$ set to $[0.1, 0.9]$, the range of threshold value $\lambda$ set to $[0.1, 0.9]$, and the search step set to $0.1$. The final parameter settings were determined by picking the ones that can lead to discovery results with the best conformance values.

Each discovered model was evaluated using the proposed fitness and precision measures, and the F1-score (i.e., harmonic mean) of the two measures was also calculated as an overall assessment for model conformance. Moreover, local diagnostics were performed on selected models with low precision and F1-score. %unexpected evaluation results.

All the experiments were conducted on a Linux system hosted on a machine with 3.5GHz Intel Xeon E5-1650 CPU and 15GB RAM, running Anaconda Python 4.7.10.

%To realize such design requires intensive human assistance in both the configuration and conduct of experiments. To address the challenge we automated the experiments by applying the principles of scientific workflow.
%A scientific workflow is usually visualized as a directed graph, where nodes representing computational steps are connected by edges, which represent the flow of data as the output of one step and the input of the subsequent. 
%We utilized the modular feature of OrgMiner to build such a workflow with the aid of a graph visualization software\footnote{Gephi: The Open Graph Viz Platform. (\url{https://gephi.org/})} and scripting.
%In this way, the configuration of experiments can be done in a form of visual programming, and the subsequent executions can be automated through scripting. The use of scientific workflow reduces the need for human attendance, while also simplifies the process of sharing the design with other researchers who wish to replicate or extend the experiments in future work.

\subsection{Results and Analysis}

This subsection presents our results and findings from the experiments.
A total number of $24$ organizational models were discovered and evaluated.
Table~\ref{tab:results_best} reports the model with the best conformance for each event log, respectively.

\begin{table}[h!]
\caption{Models discovered from the experiment dataset with the best conformance results: discovery configuration, model statistics, and model conformance.} \label{tab:results_best}
\centering
\begin{threeparttable}
    \begin{scriptsize}       
    \begin{tabular}{c|c|c|c|cc|ccc}
        \hline
        ~ \multirow{2}{*}{\bf Log} & \multicolumn{3}{c|}{\multirow{2}{*}{\bf Configuration}} & \multicolumn{2}{c|}{\bf Model statistics} & \multicolumn{3}{c}{\bf Conformance} \\
        \cline{5-9}
        & \multicolumn{3}{c|}{} & ~ \bf \#modes ~ & ~ \bf \#groups ~ & ~ \bf f. ~ & ~  \bf p. ~ & ~ \bf F1 ~ \\
        \hline
        \multirow{2}{*}{{WABO}} & CAT & \multirow{2}{*}{MOC} & \multirow{2}{*}{OS} & \multirow{2}{*}{$307$} & \multirow{2}{*}{$9$} & \multirow{2}{*}{$.876$} & \multirow{2}{*}{$.577$} & \multirow{2}{*}{$.696$} \\
                                 & (tc) & & & & & & & \\
        \hline
        \multirow{2}{*}{{BPIC17}} & CAT & \multirow{2}{*}{AHC} & \multirow{2}{*}{OS} & \multirow{2}{*}{$1884$} & \multirow{2}{*}{$10$} & \multirow{2}{*}{$.831$} & \multirow{2}{*}{$.641$} & \multirow{2}{*}{$.724$} \\
                                & (tc) & & & & & & & \\
        \hline
    \end{tabular}
    
    \begin{tablenotes}
        \item[1] Configuration: CAT = \textsf{CT+AT+TT}, tc = \textsf{trace clustering}; OS = \textsf{OverallScore}.
        \item[2] Conformance measures: f. = Fitness, p. = Precision, F1 = F1-score.
    \end{tablenotes}
    \end{scriptsize}
\end{threeparttable}
\end{table}

In the following, we first present a comparative analysis of the models obtained from using different settings for model discovery, in order to investigate how different selections of methods impact the quality of the discovered models in terms of global conformance measures. We then show how the problematic models (those with poor quality) can be examined by exploiting the proposed local diagnostic measures.

\subsubsection{Global Conformance Analysis}
To conduct the comparative analysis, models with the best conformance as shown in Table~\ref{tab:results_best} were used as baselines. The comparative analysis was performed phase by phase. For each of the three phases in the discovery approach, we compared the fitness and precision between models obtained from using various alternative methods in that phase, while keeping the selected methods in the other two phases the same as those in the baseline models. 
%the selection of methods for other phases with reference to the selected baseline models.

\paragraph{Execution mode learning} (Table~\ref{tab:results_cmp_mode_learning})
The selection of execution modes determines a way that partitions the event log data.
The conformance of the model produced from applying method \textsf{ATonly} has a higher fitness with respect to the baseline. This could be explained by the fact that applying \textsf{ATonly} leads to a relatively more coarse collection of execution modes, making the same amount of events correspond to a much smaller number of execution modes ($307$ \textit{vs.}\ $27$ and $1884$ \textit{vs.}\ $24$ modes), and therefore increases the number of events that could be fitted by correctly assigning an execution mode to a group.
In contrast, this has a negative impact on producing a precise model as the number of candidate resources associated with an event increased, which explains the observation that the precision values of models mined using \textsf{ATonly} are worse.

On the other hand, comparing models generated from using two different strategies for implementing \textsf{CT+AT+TT}, we observe that the one produced from selecting a single case attribute to derive case types (\textsf{CT+AT+TT (case attribute)}), has comparatively lower values in both fitness and precision. The possible reason is that, in terms of classifying business cases to build case types, the use of a trace clustering technique may lead to more reasonable output, since it takes into account comprehensive information from an event log instead of distinguishing cases trivially based on a single case attribute. As a result, execution modes derived from \textsf{CT+AT+TT (trace clustering)} may better capture the similarity among events at the case dimension, hence lead to organizational models with better quality.

\begin{table}[tb]
\caption{Comparison among models resulting from applying different execution mode learning methods: \textsf{CT+AT+TT (trace clustering)}, \textsf{ATonly}, and \textsf{CT+AT+TT (case attribute)}.} \label{tab:results_cmp_mode_learning}
\centering
\begin{threeparttable}
    \begin{scriptsize}       
        \begin{tabular}{c|c|c|c|cc|ccc}
            \hline
            ~ \multirow{2}{*}{\bf Log} & \multicolumn{3}{c|}{\multirow{2}{*}{\bf Configuration}} & \multicolumn{2}{c|}{\bf Model statistics} & \multicolumn{3}{c}{\bf Conformance} \\
            \cline{5-9}
            & \multicolumn{3}{c|}{} & ~ \bf \#modes ~ & ~ \bf \#groups ~ & ~ \bf f. ~ & ~  \bf p. ~ & ~ \bf F1 ~ \\
            \hline
            \multirow{6}{*}{{WABO}} & AT & \multirow{6}{*}{MOC} & \multirow{6}{*}{OS} & \multirow{2}{*}{$27$} & \multirow{2}{*}{$9$} & \multirow{2}{*}{$.947$} & \multirow{2}{*}{$.407$} & \multirow{2}{*}{$.569$} \\
            & only & & & & & & & \\
            \cline{2-2}\cline{5-9}
            & CAT & & & \multirow{2}{*}{$263$} & \multirow{2}{*}{$9$} & \multirow{2}{*}{$.843$} & \multirow{2}{*}{$.448$} & \multirow{2}{*}{$.586$} \\
            & (ca) & & & & & & & \\
            \cline{2-2}\cline{5-9}
            & CAT & & & \multirow{2}{*}{\underline{$307$}} & \multirow{2}{*}{\underline{$9$}} & \multirow{2}{*}{\underline{$.876$}} & \multirow{2}{*}{\underline{$.577$}} & \multirow{2}{*}{\underline{$.696$}} \\
            & (tc) & & & & & & & \\
            \hline
            \multirow{6}{*}{{BPIC17}} & AT & \multirow{6}{*}{AHC} & \multirow{6}{*}{OS} & \multirow{2}{*}{$24$} & \multirow{2}{*}{$10$} & \multirow{2}{*}{$.923$} & \multirow{2}{*}{$.530$} & \multirow{2}{*}{$.673$} \\
            & only & & & & & & & \\
            \cline{2-2}\cline{5-9}
            & CAT & & & \multirow{2}{*}{$2020$} & \multirow{2}{*}{$10$} & \multirow{2}{*}{$.810$} & \multirow{2}{*}{$.629$} & \multirow{2}{*}{$.708$} \\
            & (ca) & & & & & & & \\
            \cline{2-2}\cline{5-9}
            & CAT & & & \multirow{2}{*}{\underline{$1884$}} & \multirow{2}{*}{\underline{$10$}} & \multirow{2}{*}{\underline{$.831$}} & \multirow{2}{*}{\underline{$.641$}} & \multirow{2}{*}{\underline{$.724$}} \\
            & (tc) & & & & & & & \\
            \hline
        \end{tabular}
        
        \begin{tablenotes}
            \item[1] Configuration: CAT = \textsf{CT+AT+TT}, tc = \textsf{trace clustering}, ca = \textsf{case attribute}; OS = \textsf{OverallScore}.
            \item[2] Conformance measures: f. = Fitness, p. = Precision, F1 = F1-score.
        \end{tablenotes}
    \end{scriptsize}
\end{threeparttable}
\end{table}

\paragraph{Resource grouping discovery} (Table~\ref{tab:results_cmp_group_discovery})
The discovery of resource grouping identifies resource groups of which members have similar characteristics in process execution.
As for resource group discovery methods, \textsf{MOC} produced models (i.e., the baseline ones) with relatively higher fitness but lower precision, in comparison with \textsf{AHC}. This is as expected since resource groups mined using \textsf{MOC} could be overlapping whereas the others are disjoint. In a situation where groups are overlapping, a resource could be in more than one group, which therefore extends the possible range of execution modes to be assigned as capabilities in the subsequent phase for all groups it belongs to. The discovered models would thus be more flexible comparatively, hence could potentially fit more events (thus increase fitness) while also introduce an excessive number of candidate resources (thus decrease precision).

\begin{table}[tb]
    \caption{Comparison among models resulting from applying different resource grouping discovery methods: \textsf{MOC} and \textsf{AHC}.}\label{tab:results_cmp_group_discovery}
    \centering
    \begin{threeparttable}
        \begin{scriptsize}       
            \begin{tabular}{c|c|c|c|cc|ccc}
                \hline
                ~ \multirow{2}{*}{\bf Log} & \multicolumn{3}{c|}{\multirow{2}{*}{\bf Configuration}} & \multicolumn{2}{c|}{\bf Model statistics} & \multicolumn{3}{c}{\bf Conformance} \\
                \cline{5-9}
                & \multicolumn{3}{c|}{} & ~ \bf \#modes ~ & ~ \bf \#groups ~ & ~ \bf f. ~ & ~  \bf p. ~ & ~ \bf F1 ~ \\
                \hline
                \multirow{4}{*}{{WABO}} & & \multirow{2}{*}{AHC} & \multirow{4}{*}{OS} & \multirow{2}{*}{$307$} & \multirow{2}{*}{$9$} & \multirow{2}{*}{$.815$} & \multirow{2}{*}{$.595$} & \multirow{2}{*}{$.687$} \\
                & CAT & & & & & & & \\
                \cline{3-3}\cline{5-9}
                & (tc) & \multirow{2}{*}{MOC} & & \multirow{2}{*}{\underline{$307$}} & \multirow{2}{*}{\underline{$9$}} & \multirow{2}{*}{\underline{$.876$}} & \multirow{2}{*}{\underline{$.577$}} & \multirow{2}{*}{\underline{$.696$}} \\
                & & & & & & & & \\
                \hline
                \multirow{4}{*}{{BPIC17}} & & \multirow{2}{*}{AHC} & \multirow{4}{*}{OS} & 
                \multirow{2}{*}{\underline{$1884$}} & \multirow{2}{*}{\underline{$10$}} & \multirow{2}{*}{\underline{$.831$}} & \multirow{2}{*}{\underline{$.641$}} & \multirow{2}{*}{\underline{$.724$}} \\
                & CAT & & & & & & & \\
                \cline{3-3}\cline{5-9}
                & (tc) & \multirow{2}{*}{MOC} & & \multirow{2}{*}{$1884$} & \multirow{2}{*}{$8$} & \multirow{2}{*}{.957} & \multirow{2}{*}{.406} & \multirow{2}{*}{.571} \\
                & & & & & & & & \\
                \hline
            \end{tabular}
            
            \begin{tablenotes}
                \item[1] Configuration: CAT = \textsf{CT+AT+TT}, tc = \textsf{trace clustering}; OS = \textsf{OverallScore}.
                \item[2] Conformance measures: f. = Fitness, p. = Precision, F1 = F1-score.
            \end{tablenotes}
        \end{scriptsize}
    \end{threeparttable}
\end{table}

\paragraph{Execution mode assignment} (Table~\ref{tab:results_cmp_mode_assignment})
Assigning execution modes to discovered resource groups yields the final organizational models.
By using method \textsf{FullRecall} one may sacrifice much precision for perfect fitness, as redundant execution modes would be recognized as group capabilities and thus make a produced model too flexible --- in the sense that resources could be allowed to carry out an excessive number of modes. Similar to a ``flower'' model~\cite{vanderaalst2016process} in process model discovery,
%\footnote{Note that the idea behind such an organizational model is only similar to but should not be regarded as equivalent to that behind a ``flower'' model. In this context, an equivalent organizational model should allow any resource to be capable of executing all involved modes.}
such an organizational model is able to fit everything observed in a log, but may still be inappropriate in terms of modeling the actual behavior of resources in an organization underlying the log data as it lacks precision. By comparison, the baseline models resulted from applying method \textsf{OverallScore} gained much better precision while still maintaining a moderate level of fitness.

\begin{table}[tb]
    \caption{Comparison among models resulting from applying different execution mode assignment methods: \textsf{OverallScore} and \textsf{FullRecall}.}\label{tab:results_cmp_mode_assignment}
    \centering
    \begin{threeparttable}
        \begin{scriptsize}       
            \begin{tabular}{c|c|c|c|cc|ccc}
                \hline
                ~ \multirow{2}{*}{\bf Log} & \multicolumn{3}{c|}{\multirow{2}{*}{\bf Configuration}} & \multicolumn{2}{c|}{\bf Model statistics} & \multicolumn{3}{c}{\bf Conformance} \\
                \cline{5-9}
                & \multicolumn{3}{c|}{} & ~ \bf \#modes ~ & ~ \bf \#groups ~ & ~ \bf f. ~ & ~  \bf p. ~ & ~ \bf F1 ~ \\
                \hline
                \multirow{4}{*}{{WABO}} & & \multirow{4}{*}{MOC} & \multirow{2}{*}{FR} & \multirow{2}{*}{$307$} & \multirow{2}{*}{$9$} & \multirow{2}{*}{$1.000$} & \multirow{2}{*}{$.067$} & \multirow{2}{*}{$.125$} \\
                & CAT & & & & & & & \\
                \cline{4-4}\cline{5-9}
                & (tc) & & \multirow{2}{*}{OS} & \multirow{2}{*}{\underline{$307$}} & \multirow{2}{*}{\underline{$9$}} & \multirow{2}{*}{\underline{$.876$}} & \multirow{2}{*}{\underline{$.577$}} & \multirow{2}{*}{\underline{$.696$}} \\
                & & & & & & & & \\
                \hline
                \multirow{4}{*}{{BPIC17}} & & \multirow{4}{*}{AHC} & \multirow{2}{*}{FR} & \multirow{2}{*}{$1884$} & \multirow{2}{*}{$10$} & \multirow{2}{*}{$1.000$} & \multirow{2}{*}{$.169$} & \multirow{2}{*}{$.290$} \\
                & CAT & & & & & & & \\
                \cline{4-4}\cline{5-9}
                & (tc) & & \multirow{2}{*}{OS} & \multirow{2}{*}{\underline{$1884$}} & \multirow{2}{*}{\underline{$10$}} & \multirow{2}{*}{\underline{$.831$}} & \multirow{2}{*}{\underline{$.641$}} & \multirow{2}{*}{\underline{$.724$}} \\
                & & & & & & & & \\
                \hline
            \end{tabular}
            
            \begin{tablenotes}
                \item[1] Configuration: CAT = \textsf{CT+AT+TT}, tc = \textsf{trace clustering}; OS = \textsf{OverallScore}, FR = \textsf{FullRecall}.
                \item[2] Conformance measures: f. = Fitness, p. = Precision, F1 = F1-score.
            \end{tablenotes}
        \end{scriptsize}
    \end{threeparttable}
\end{table}

The analysis of comparing the conformance of organizational models obtained from various settings showcases a useful application of the proposed conformance checking measures: evaluating discovery algorithms through measuring the quality of discovered organizational models. This extends the literature and contributes to future work on organizational model mining, as the effectiveness of a developed discovery method could be evaluated against the input event log.
On the other hand, it could also benefit any application of discovering organizational models from event logs by serving as a reference for configuring and adjusting the method in use, in order to obtain results of satisfying quality.

\subsubsection{Local Diagnostic Analysis}
From the presented experimental results, one may notice that some methods led to models suffering unexpectedly low conformance due to the lack of precision. In addition to investigating the cause from the perspective of the discovery approach, we also consider the discovered models ``as-is'' models, and applied local diagnostic measures to examine the potential reasons behind non-conformance.
We took as an example an organizational model derived from log {WABO} during the experiments, since it has a poor precision and comparatively the lowest F1-scores among all models discovered\footnote{
    Note that these refer to the complete set of results from model discovery rather than the ones presented for the comparative analysis on global conformance.
}, i.e., the model resulting from applying \textsf{CT+AT+TT (case attribute)}-\textsf{MOC}-\textsf{FullRecall}, with fitness = $1.0$, precision $= 0.036$ and F1-score $= 0.069$.

The definition of precision (Def.~\ref{def:precision}) suggests that an organizational model is likely to be imprecise when it allows an exaggerated number of candidate resources for an event (especially when the candidates involve those who have never actually executed the event). 
From a model point of view, this may be reflected as resource groups having excessive capabilities allowed with respect to event log data. 

In order to reveal such defects in the problematic model, we calculated the local diagnostic measures for each ``group-mode'' pair allowed in the model.
We then combined all the obtained scores into a list and sorted all ``group-mode'' pairs, which subsequently enabled us to locate group capabilities which could be causing the non-conformance between the model and the event log. For illustration purposes, we selected part of the diagnostic results for our discussion.

\begin{landscape}
\begin{figure*}[t]
    \centering
    \includegraphics[width=\linewidth]{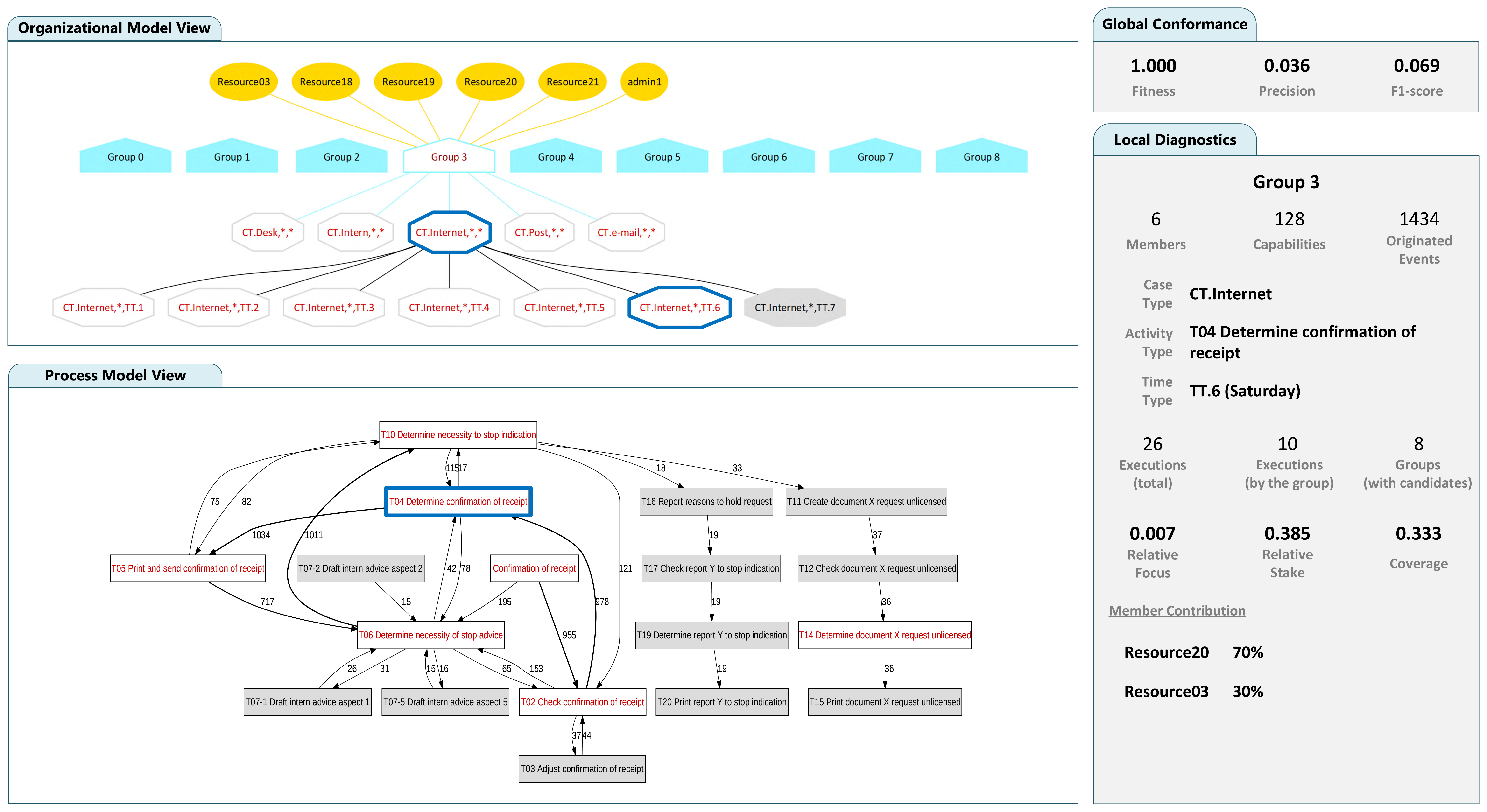}
    \caption{Local diagnostics example: detecting non-conformance on ``Group 3'' and its capable execution mode (``CT.Internet'', ``T04 Determine confirmation of receipt'', ``TT.6'') of the selected organizational model. Note that the activity types of execution modes are projected onto a corresponding process model discovered as a Directly-Follows Graph where the most frequent edges are shown.} \label{fig:results_diagnostic_wabo1}
\end{figure*}
\end{landscape}

Fig.~\ref{fig:results_diagnostic_wabo1} shows a resource group in the model with $6$ members, ``Group 3'', which is capable of carrying out execution modes across all types of cases.
According to the model, this group is allowed to carry out a total of $133$ execution modes, yet the diagnostic results show only $46$ of them have a group coverage of over $0.5$, implying that in the actual process execution --- as recorded in the event log --- the majority of capabilities ($87$ of $133$) of this resource group have been carried out by merely one or two members rather than been shared across the whole group. For instance, as shown in Fig.~\ref{fig:results_diagnostic_wabo1}, execution mode (``CT.Internet'', ``T04 Determine confirmation of receipt'', ``TT.6'')\footnote{
    TT.1, TT.2, \ldots TT.7 denote the time types corresponded to the seven weekdays respectively used in the current experimental settings (see Sect.~\ref{sec:evaluation/setup}).
} was carried out by two of the group members only, ``Resource~03'' and ``Resource~20'', whereas the other four members have never participated in the work as observed in the log data.

\begin{landscape}
\begin{figure*}[t]
    \centering
    \includegraphics[width=\linewidth]{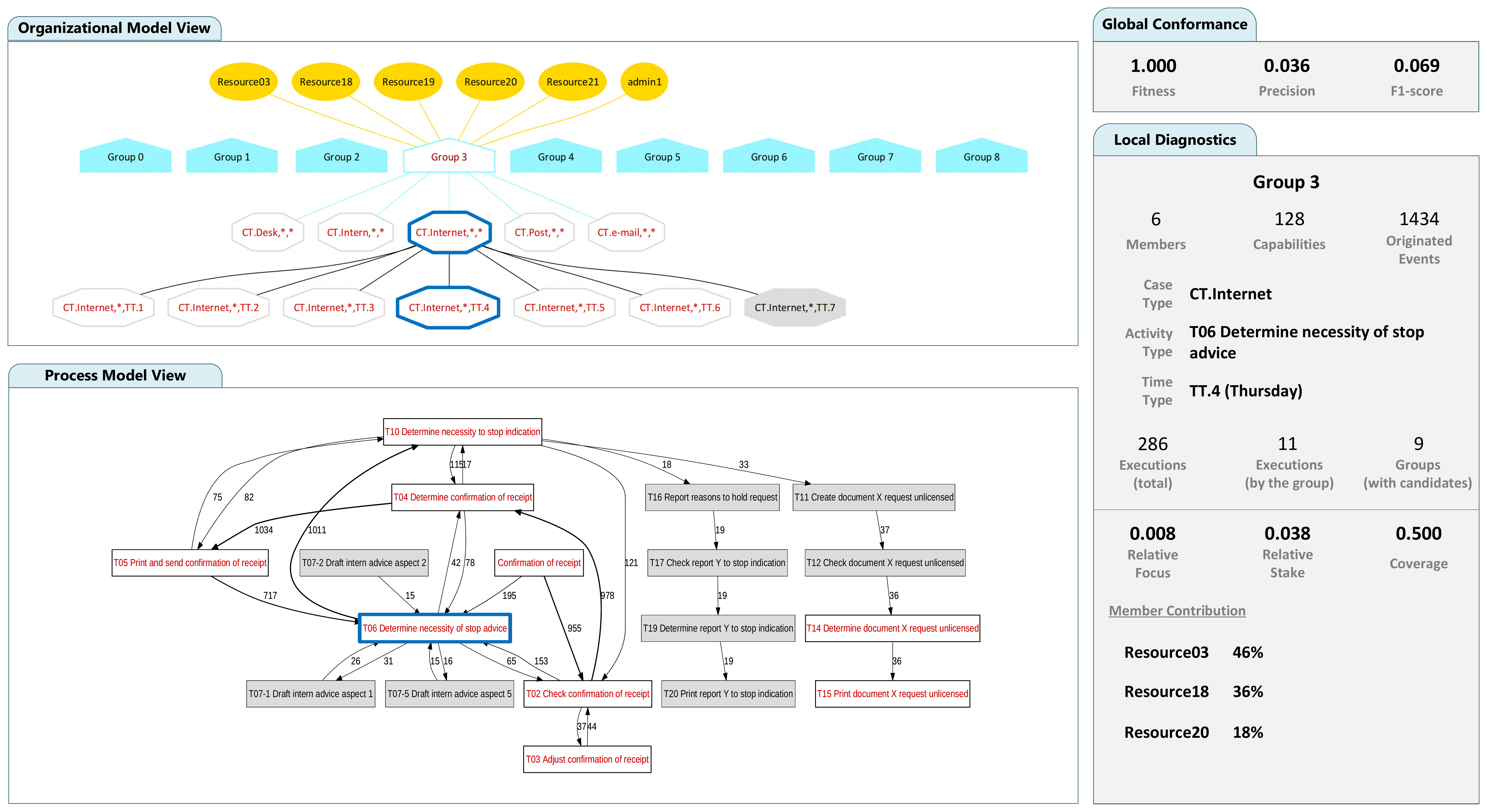}
    \caption{Local diagnostics example: detecting non-conformance on ``Group 3'' and its capable execution mode  (``CT.Internet'', ``T06 Determine necessity of stop advice'', ``TT.4'') of the selected organizational model. Note that the activity types of execution modes are projected onto a corresponding process model discovered as a Directly-Follows Graph where the most frequent edges are shown.} \label{fig:results_diagnostic_wabo2-1}
\end{figure*}
\end{landscape}

\begin{landscape}
\begin{figure*}[t]
    \centering
    \includegraphics[width=\linewidth]{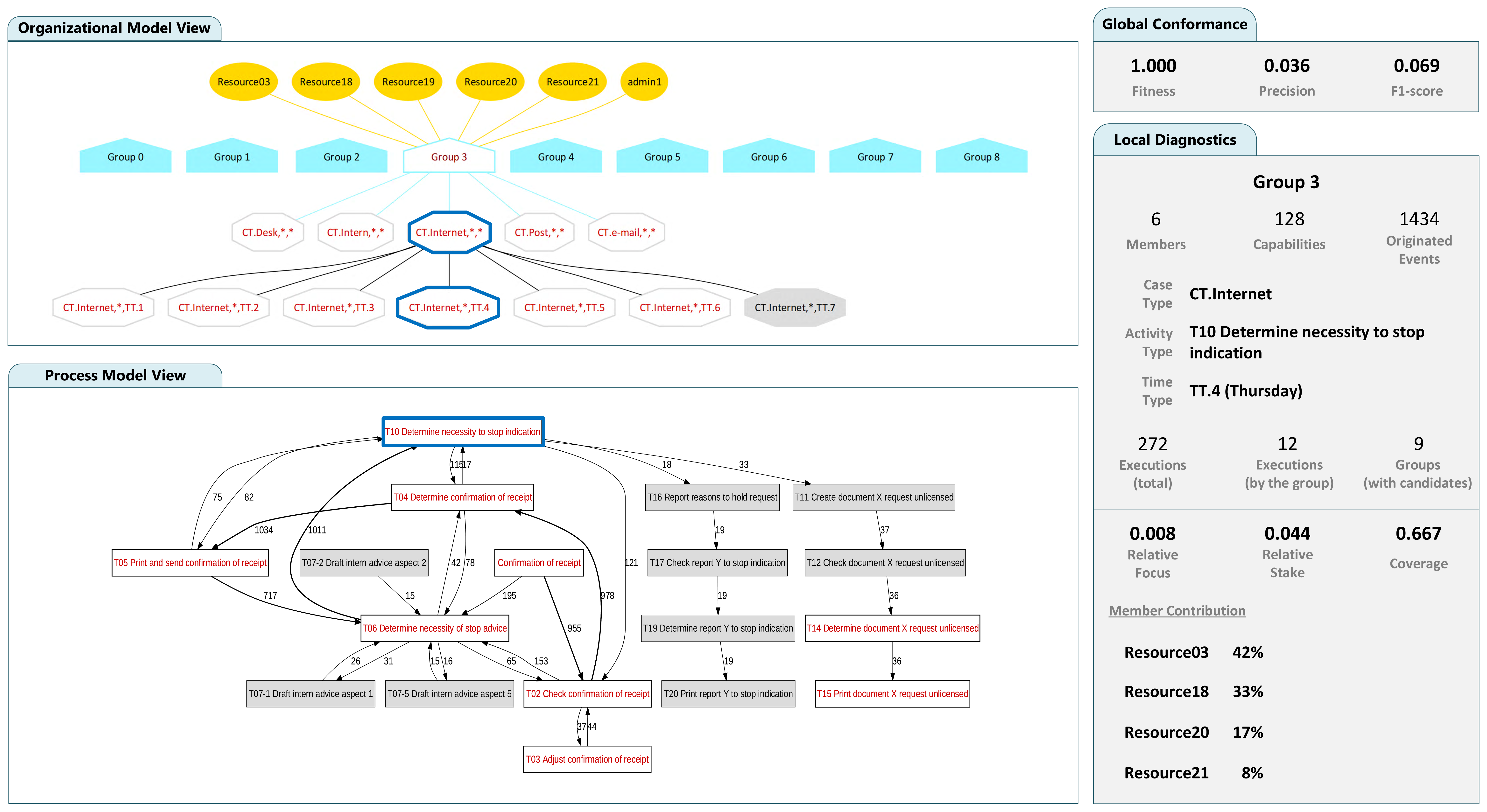}
    \caption{Local diagnostics example: detecting non-conformance on ``Group 3'' and its capable execution mode (``CT.Internet'', ``T10 Determine necessity of stop indication'', ``TT.4'') of the selected organizational model. Note that the activity types of execution modes are projected onto a corresponding process model discovered as a Directly-Follows Graph where the most frequent edges are shown.} \label{fig:results_diagnostic_wabo2-2}
\end{figure*}
\end{landscape}

On the other hand, the results of examining group relative stake scores also provided us interesting findings (Fig.~\ref{fig:results_diagnostic_wabo2-1} and Fig.~\ref{fig:results_diagnostic_wabo2-2}). For example, there are two execution modes accounted as the group's capabilities, i.e., (``CT.Internet'', ``T06 Determine necessity of stop advice'', ``TT.4'') and (``CT.Internet'', ``T10 Determine necessity of stop indication'', ``TT.4''), which were executed $286$ and $272$ times according to the log, however only less than $5\%$ of the total work on these execution modes was actually committed by members from the group (as revealed by group relative stake).

The above analysis shows an example of using local diagnostics to locate the imprecision of an organizational model with respect to a particular resource group within the model. For the example group selected, we were able to recognize group capabilities (as execution modes) that are allowed in the model but seem irrelevant according to actual process execution: either carried out by only a small proportion of the group (low group coverage), or not involving the group as a major contributor (low group relative stake). 
The analysis can also be conducted for other resource groups to detect imprecision at the local level, and the combined results may lead to a full picture reflecting where the organizational model does not align with the log data.
Depending on the intention of the model, this could inform different possible actions for improvement. In this case, as the diagnosed model was derived from applying a model discovery approach, a potential further step could be improving the approach to avoid assigning those less related execution modes as capabilities of groups, thus to discover more precise models.

\section{Conclusion} \label{sec:conclusion}
This paper contributes to organizational model mining, a subfield of process mining, by proposing a framework based on a novel definition of organizational models in the context of business processes. Such an organizational model describes the linkage from human resources via resource groups to the activities, case types, and time periods of a business process. Compared to the state-of-the-art, the presented definition of organizational models covers not only the structural (group membership) but also the behavioral (group capabilities) aspects of resource grouping, bridging the analysis of human resources and their organizational structure with process analytics.
Another novelty of the reported work lies in the proposal of conformance checking concepts built around the defined organizational models, which enables one to evaluate a model against event log data in terms of both measuring the degree (by global conformance checking) and examining the reasons (through local diagnostics) for commonalities or discrepancies between the two. This distinguishes the current work from the literature as it provides a rigorous means to analyze organizational models with respect to event logs, which not only satisfies but can also extend beyond the purpose of assessing the effectiveness of a discovery method.

We have demonstrated the feasibility of the framework by presenting a model discovery approach and conducting experiments on real-life event logs to evaluate the discovered models. One may notice from the approach how some of the previous work on organizational model mining can be incorporated into the current framework. On the other hand, the reported experimental results and findings have showcased part of the many potential applications of the proposed conformance utilities, including comparative analysis of model discovery methods, and deviation analysis on organizational models with respect to the actual behavior observed in event logs.

Our organizational model mining framework opens up many interesting questions to be explored in future work. All three standard types of process mining can be addressed for organizational models within the framework. For instance, (1) developing more effective methods in order to discover organization models from event logs with more satisfying conformance, (2) extending the current set of conformance checking measures to consider dimensions beyond model fitness and precision, and (3) repairing organizational models according to local diagnostic outcomes.

\section*{Acknowledgments}
The reported research is part of a PhD project supported by an Australian Government Research Training Program (RTP) Scholarship.
We thank the Alexander von Humboldt (AvH) Stiftung for supporting our research.
This work is also supported by the National Natural Science Foundation of China (NSFC) under Grants No.~61972427 and U1911205.

%\section*{References}
\bibliography{orgmining}

\end{document}